\documentclass[twocolumn,aps,pra,superscriptaddress,showpacs]{revtex4-1}

\usepackage{amssymb}
\usepackage{amsmath}
\usepackage{mathtools}
\usepackage{graphicx}
\usepackage{bm}
\usepackage[english]{babel}   
\usepackage{hyperref}
\hypersetup{%
   pdfpagemode=UseNone, 
   pdfstartpage=1,
   pdfmenubar=true,
   pdftoolbar=true,
   colorlinks = true,
   linkcolor=blue,
   citecolor=blue,
   urlcolor=blue,
   bookmarksopen=false
 }

\binoppenalty=\maxdimen
\relpenalty=\maxdimen

\begin{document}

\title{Vortex dynamics in coherently coupled Bose-Einstein condensates}
\author {Luca Calderaro}
\affiliation {Dipartimento di Ingegneria dell'Informazione, Universit\`a di Padova, 35131 Padova, Italy}
\author {Alexander L.\ Fetter}
\email{fetter@stanford.edu }
\affiliation {Departments of Physics and Applied Physics, Stanford University,
   Stanford, CA 94305-4045, USA}
\author {Pietro Massignan}
\email{pietro.massignan@icfo.es}
\affiliation{ICFO -- Institut de Ciencies Fotoniques, The Barcelona Institute of Science and Technology, 08860 Castelldefels (Barcelona), Spain}
\author {Peter Wittek}
\affiliation{ICFO -- Institut de Ciencies Fotoniques, The Barcelona Institute of Science and Technology, 08860 Castelldefels (Barcelona), Spain}
\affiliation {University of Bor{\aa}s, 50190 Bor{\aa}s, Sweden}

\date{\today}

\begin {abstract}
In classical hydrodynamics with uniform density, vortices move with the local fluid velocity.  This description is rewritten in terms of forces arising from the interaction with other vortices. Two such positive straight vortices experience a repulsive interaction and precess in a positive (anticlockwise) sense around their common centroid. A similar picture applies to vortices in a two-component two-dimensional uniform Bose-Einstein condensate (BEC) coherently coupled through rf Rabi fields.  Unlike the classical case, however, the rf Rabi coupling induces an attractive interaction and two such vortices with positive signs now rotate in the negative (clockwise) sense. Pairs of counter-rotating vortices are instead found to translate with uniform velocity perpendicular to the line joining their cores. This picture is extended to a single vortex in a two-component trapped BEC. Although two uniform vortex-free components experience familiar Rabi oscillations of particle-number difference, such behavior is absent for a vortex in one component because of the nonuniform vortex phase. Instead the coherent Rabi coupling induces a periodic vorticity transfer between the two components.
\end{abstract}
\maketitle

\section{Introduction}\label{sec1}

Onsager and Feynman revolutionized superfluid physics with the concept of quantized vortex lines.  Originally, this idea was introduced to describe superfluid $^4$He, but it also applied to the more recent ultracold atomic Bose-Einstein condensates (BECs).  Initial vortex research emphasized the equilibrium configurations, for example in rotating superfluid BECs where imaging an expanded condensate provided direct visualization of the vortex arrays.

In certain cases for atomic BECs, however, the dynamics of one or two vortices is not only calculable but also observable experimentally in real time, providing a rare opportunity to study such time-dependent phenomena.  Note that the analogous vortex dynamics in superfluid $^4$He is largely inaccessible owing to the very small vortex core.  Here we analyze the effect of coherent rf Rabi coupling on the dynamics of one or two vortices in a two-component BEC mixture.

The physics of two coupled Bose-Einstein condensates has been of great  interest since the early JILA experiments using two hyperfine states of $^{87}$Rb~\cite{Hall98}.  Initially, these coupled condensates had the usual mean-field interactions, in which case the typical  Gross-Pitaevskii equation contains two interaction terms  proportional to the two local particle densities $n_1$ and $n_2$.  Correspondingly, the  interaction  energy density is ${\cal E}_{\rm int} =\frac{1}{2} \sum_{ij= 1,2}\,g_{ij}\, n_i n_j$, where $n_j =|\psi_j|^2$ is the condensate density for component $j$,  and $g_{ij}$ is a set of interaction parameters. Since  this interaction energy depends solely on the densities,  it  carries no information on the relative phase of the two condensates.  

Subsequently, the JILA group  added  coherent  rf Rabi fields involving direct  linear off-diagonal coupling of the two order parameters~\cite{Matt98,Matt99,Ande00}.  In the time-dependent GP equation for (say) $\psi_1$, there is a  term with $\psi_2$  proportional to the Rabi frequency $\Omega$,  which is related to the strength of the rf coupling fields.  The corresponding  coherent rf Rabi interaction energy density  ${\cal E}_\Omega = -\frac{1}{2}\hbar\Omega\, \Psi^\dagger \sigma_x\Psi = -\frac{1}{2}\hbar\Omega (\psi^\dagger_1\psi_2 + \psi^\dagger_2\psi_1) $ is very different  from the more familiar mean-field  form given above. As a result, the two components now form a coupled two-level system with dynamics analogous to coherent motion on a Bloch sphere.

In  2002~\cite{Son02}, Son and Stephanov pointed out the crucial role of such coherent rf Rabi coupling, emphasizing the presence of a narrow domain wall between two vortices, whose dynamics closely mimics string-breaking processes in quantum chromodynamics.   With the  density-phase representation of the condensate order parameters  $\psi_j = \sqrt{n_j}\,e^{iS_j}$,  the coherent coupling energy density becomes  ${\cal E}_\Omega = -\hbar\Omega\sqrt{n_1n_2}\cos(S_1-S_2)$, involving the phase difference between the two condensates.  Note that this long-wavelength  rf coupling is spatially uniform, in contrast to the finite-wavelength Raman coupling introduced by Spielman~\cite{Spie09,Lin09}, where the spatial dependence of the coupling term is significant.

The Lagrangian density is ${\cal L} = {\cal T} - {\cal E}_{\rm GP}$, where ${\cal T} =\frac{1}{2}i\hbar [\Psi^\dagger\partial _t\Psi -(\partial_t\Psi^\dagger) \Psi]$ and the remaining term is the usual GP energy density functional, including the kinetic energy, the trap energy,  the interaction energy and the Rabi coupling energy.  Expressed in terms of number density and phase, the Lagrangian density 
becomes
\begin{align}\label{L2}
{\cal L} =& \sum_{i=1,2}\left[-\hbar n_i\dot{S}_i - \frac{\hbar^2}{2M} \left(\bm\nabla \sqrt{n_i}\right)^2  -\frac{\hbar^2}{2M}n_i\left(\bm \nabla S_i\right)^2 \right]\nonumber\\
  &+ \hbar\Omega\sqrt{n_1n_2}\cos(S_1-S_2) -\frac{1}{2} gn^2 +\delta g_{12}n_1n_2,
\end{align}
where $n=n_1+n_2$ is the total number density.  Here and throughout, the trap is omitted (except for Secs.\ \ref{coherenttrap} and \ref{JosephsonDynamics}) and  this simple model assumes  interaction constants $g_{11}=g_{22} =g$ and $\delta g_{12} = g- g_{12} >0$.  These parameters are appropriate for $^{87}$Rb and imply that the  uniform system does not phase separate.   This form of the Lagrangian density is useful for studying the dynamics of vortices in coherently coupled BECs.
Much of the present analysis will focus on a tightly confined effectively two-dimensional condensate, in which case $n_j$ represents a two-dimensional particle density with dimension of an inverse area and the corresponding coupling constants are renormalized by the tight axial harmonic trapping potential  $g\to g_{\rm 2D} = g/(d_z\sqrt{2\pi})$, where $d_z =\sqrt{\hbar/(M\omega_z)}$ is the axial oscillator length.  For simplicity, we will use $g$ to denote the two-dimensional coupling constant with units of energy $\times$ area.  Hence $gM/\hbar^2$ is dimensionless.

Section~\ref{dyn} briefly reviews the dynamics of classical rectilinear vortices in a single-component fluid. Section~\ref{force} then rewrites the dynamical equations in terms of forces arising from intervortex potentials;  it also  derives the same vortex dynamics for a one-component dilute BEC from a variational Lagrangian formalism.   Section~\ref{domain} summarizes the essential features of the coherent coupling in a two-component BEC from~\cite{Son02}, focusing on the domain wall of relative phase.  These various features combine in Section~\ref{2BEC} to describe the dynamics of two vortices in coherently coupled uniform BECs with one vortex in each component.  Section~\ref{coherenttrap} studies instead the dynamics for a single vortex in a  trapped condensate, where the coherent coupling induces periodic transfer of vorticity between the two condensates.
Section~\ref{JosephsonDynamics} then investigates the Josephson-like dynamics of the coherent transfer of population between two coherently coupled condensates.  In the absence of a vortex, the population difference exhibits  familiar Rabi oscillation~\cite{Matt98}. When a vortex is present in one condensate, however, the lack of overall global phase leads to a cancelation,  and instead the vorticity transfers periodically between the two components with no coherent population transfer, in analogy with similar results for coherently coupled annular condensates~\cite{Gall15}.

\section{Vortex Dynamics in Classical Hydrodynamics}\label{dyn}

In thinking about vortex dynamics in two-dimensional BECs, it is simplest to start from classical incompressible hydrodynamics and focus on a set of point vortices at   $\bm r_j$.  Each vortex generates its own circulating velocity field
\begin{equation}\label{vj}
\bm v_j(\bm r) = q_j\, \frac{\hbar}{M}\,\frac{\hat{\bm z}\times (\bm r-\bm r_j)}{|\bm r-\bm r_j|^2},
\end{equation}
where $q_j = \pm 1$ characterizes the sense of circulation, which is quantized in units of  $2\pi\hbar/M$ (alternatively, the velocity is $\hbar/M$ times the gradient of the phase $S_j$).  A given vortex at $\bm r_i$ has a translational velocity
\begin{equation}\label{vortexdyn}
\dot{\bm r}_i = \sum_{j\neq i} \bm v_j(\bm r_i)= \sum_{j\neq i} q_j\frac{\hbar}{M}\,\frac{\hat{\bm z}\times (\bm r_i-\bm r_j)}{|\bm r_i-\bm r_j|^2}
\end{equation}
 equal to the total velocity at its position induced by all the other vortices (and, in principle, any additional imposed flow).  

It is helpful to focus on two vortices at $\bm r_1$ and $\bm r_2$ separated by a distance $r_{12} $.
Their dynamical equations lead to the expected dynamics
\begin{equation}\label{1}
\dot {\bm r}_1 =q_2 \frac{\hbar}{M}\,\hat{\bm z}\times\frac{\bm r_1-\bm r_2}{|\bm r_1-\bm r_2|^2},
\end{equation}
so that vortex 1 moves with the velocity induced at its location by vortex 2.
Similarly,
\begin{equation}\label{2}
\dot {\bm r}_2 = q_1\frac{\hbar}{M}\,\hat{\bm z}\times\frac{\bm r_2-\bm r_1}{|\bm r_2-\bm r_1|^2}.
\end{equation}
If they have the same circulation with $q_1q_2 = 1$,
they rotate at fixed $r_{12}$ around their joint center at an instantaneous linear speed $\hbar/(Mr_{12})$ with a sense determined by their individual circulations [equivalently, the angular velocity around the common center is $2\hbar/(Mr_{12}^2)$].  If they have opposite circulations $q_1q_2=-1$, they  are called a vortex pair or a vortex dipole and move uniformly at fixed $r_{12}$ with translational velocity $\hbar/(Mr_{12})$ in the direction of the flow between them.  The Arizona group~\cite{Neel10} has created such vortex dipoles   reproducibly in disk-shaped BECs and followed their dynamical trajectories.  The finite boundaries significantly affect the motion, and the experiment 
 observed one full cycle of the  vortex dipole orbits.

As seen below, the energy of two vortices in an unbounded medium depends only on the distance between them, so that both these dynamical motions maintain the total energy.  This behavior reflects the lack of any dissipative mechanism in classical hydrodynamics.

In a uniform dilute Bose gas obeying the Gross-Pitaevskii (GP) equation, the same result holds as long as the vortices are well separated relative to the healing length $\xi = \hbar/\sqrt{2Mng}$, which characterizes the vortex core radius.  Reference~\cite{Fett66}  proved this result by direct examination of the time-dependent GP equation, assuming that the time dependence arises solely from the rigid motion of the vortices. This result is not surprising, for the time-dependent GP equation implies both the usual conservation of particles and a Bernoulli equation for isentropic compressible flow;  these two suffice to describe classical inviscid hydrodynamics, including vortex motion~\cite{Fett09}.

\section{Vortex Response to Applied Force}\label{force}

From one perspective, Eq.~(\ref{vortexdyn}) wholly suffices to describe the motion of  point vortices in a uniform two-dimensional fluid,  but it is instructive to generalize and think of effective forces.  Note the simple identity~\cite{Kim04}
 \begin{equation}\label{ln}
\frac{\bm r}{r^2}  = -\bm\nabla\ln \left(\frac{1}{r}\right),
\end{equation}
where $\ln(1/r) $ is essentially the Coulomb Green's function in two dimensions.  This approach is equivalent to the use of a stream function instead of a velocity potential.  Define
\begin{equation}\label{Vij}
\tilde V_{ij}(r) = 2\pi\hbar n\,q_iq_j \frac{\hbar}{M} \ln\left(\frac{1}{r}\right) = 2\pi \hbar n\,V_{ij}(r),
\end{equation}
where $V_{ij}(r) =   q_iq_j\hbar \ln(1/r)/M$ omits  the  dimensional factor $2\pi \hbar  n$.
Here, $\tilde V_{ij}(r) $   is the interaction energy  between two point vortices in two dimensions.  Note that for two vortices with the same sign $q_1q_2 = 1$, the interaction is repulsive and diverges  to $\infty$ as $r\to 0$, whereas for two vortices with opposite sign $q_1q_2 = -1$, it is attractive and diverges to $-\infty$ as $r\to 0$.

In particular,  again focus on two vortices  in a one-component fluid, in which case the equations of vortex motion now become
\begin{eqnarray}\label{force12}
\nonumber q_1\dot{\bm r}_1 &=& -\hat{\bm z}\times \bm\nabla_1 V_{12}(r_{12}), \\
q_2 \dot{\bm r}_2 &=& -\hat{\bm z}\times \bm\nabla_2 V_{12}(r_{12}).
\end{eqnarray}
Apart from an overall factor, the quantity  $-\bm \nabla_1V_{12}(r_{12})$ is effectively the force $\bm F_1$ that vortex 2 exerts on vortex 1.  Hence Eq.~(\ref{force12}) assumes the simple and physical form
\begin{eqnarray}\label{Magnus}
\nonumber q_1\dot{\bm r}_1 &=& \hat{\bm z}\times\bm F_1\\
q_2 \dot{\bm r}_2 &=& \hat{\bm z}\times \bm F_2  = -\hat{\bm z} \times \bm F_1,
\end{eqnarray}
where $\bm F_1=-\bm F_2$,  since they arise from a central potential.
It says that each vortex moves perpendicular to  the force $\bm F$ on it, in a direction determined by $q_j\,\hat{\bm z}\times \bm F_j$.  Such behavior is often called the Magnus effect.

By inspection,  the vector quantity $q_1\bm r_1 + q_2\bm r_2$ is conserved. Also, Eqs.~(\ref{1}) and (\ref{2}) show that the relative vector $\bm r_{12} = \bm r_1-\bm r_2$ obeys the dynamical equation
\begin{equation}\label{dotr12}
\dot{\bm r}_{12} = \frac{\hbar}{Mr_{12}^2} \left(q_1+q_2\right) \,\hat{\bm z}\times \bm r_{12}.
\end{equation}
   If $q_1=q_2 = 1$, then the two vortices precess around each other at fixed separation with an angular velocity $2\hbar/(Mr_{12}^2)$ in the positive sense, as found previously. If $q_1=- q_2=1$ (a vortex pair/vortex dipole), then $\bm r_1 -\bm r_2$ remains constant,  and  Eq.~(\ref{Magnus}) indicates that $\frac{1}{2}\left(\dot{\bm r}_1 + \dot{\bm r}_2\right) = \hat{\bm z}\times \bm F_1$, so that the center of the pair moves uniformly.

The operation $\hat{\bm z}\times$ allows these dynamical relations to be rewritten as $\bm F_j^M+ \bm F_j=0$, where
\begin{equation}\label{Magnusforce}
\bm F_j^M \equiv q_j\,\hat{\bm z}\times\dot{\bm r}_j
\end{equation}
is called the Magnus force.
In this latter form, the vector sum of all forces acting on the vortex must vanish, which thus determines the motion of the vortex.
Effectively, a vortex has  intrinsic angular momentum arising from its circulating flow and acts like a gyroscope.

For subsequent reference, it is also useful to study the behavior of two vortices in a uniform single-component two-dimensional BEC with the time-dependent Lagrangian formalism, which is equivalent to Eq.~(\ref{L2}) with only a single uniform two-dimensional density $n$ (ignoring the vortex core structure)  and phase $S$.  Assume two vortices at $\bm r_1$ and $\bm r_2$ with unit circulations $q_1$ and $q_2$ and total phase $S = S_1 + S_2$, where
\begin{equation}\label{phasej}
S_j = q_j \,{\arg}\,[(x-x_j) + i(y-y_j)]= q_j\arg(z-z_j),
\end{equation}
where ${\arg}(z) = z/|z|$ is the phase of the complex number $z=x+iy$.  Here,    $S_1$ and $S_2$  refer to distinct vortices in the same component.  Based on this form for the phase arising from  each vortex, it is not hard to find the time-dependent term in the Lagrangian
\begin{equation}\label{T}
T = \hbar \pi n \left(q_1 \hat {\bm z} \times \dot {\bm r}_1 \cdot \bm r_1 + q_2 \hat {\bm z} \times \dot {\bm  r}_2 \cdot \bm r_2\right),
\end{equation}
which is unusual in depending linearly on the coordinate and the velocity of each vortex.

 The corresponding  fluid velocity is $\bm v_1 + \bm v_2$, where $\bm v_j$ is given in (\ref{vj}),
 and the kinetic energy density  is $\frac{1}{2}Mn\left(\bm v_1 + \bm v_2\right)^2$.
 Apart from the divergent self-energy of each vortex,  the interaction energy density is ${\cal E}_{12}= Mn \,\bm v_1\cdot\bm v_2 = (q_1q_2\hbar^2n /M) \bm \nabla\ln|\bm r-\bm r_1| \cdot \bm \nabla \ln |\bm r-\bm r_2|$.  The interaction energy $E_{12}= \int d^2 r\,{\cal E}_{12}$  involves a two-dimensional  integral, which may be computed using the divergence theorem and the two-dimensional Coulomb Green's function $G_2(r) = -\ln(r)$ that  satisfies the  equation $\nabla^2 G_2(r) = -2\pi \delta^{(2)}(\bm r)$  [equivalently, $\nabla^2\ln r = 2\pi\delta^{(2)}(\bm r)$].
 As a result,
 \begin{equation}\label{E12}
 E_{12} =q_1q_2 \, \frac{2\pi\hbar^2 n}{M} \ln\frac{1}{|\bm r_1-\bm r_2|},
 \end{equation}
  which is just the interaction energy $\tilde V_{12}(r_{12})$ from Eq.~(\ref{Vij}).

Hence the total Lagrangian becomes
\begin{multline}\label{Lonecomp}
L= \hbar \pi n \left(q_1 \hat {\bm z} \times \dot {\bm r}_1 \cdot \bm r_1 + q_2 \hat {\bm z} \times \dot {\bm  r}_2 \cdot \bm r_2\right)\\
-q_1q_2\,\frac{2\pi \hbar^2 n}{M}\,\ln\frac{1}{| \bm r_1-\bm r_2 |}.
\end{multline}
Focus on vortex 1, when $\partial {L}/\partial \dot{\bm r}_1 = -\hbar \pi nq_1\hat{\bm z}\times \bm r_1$.  The Euler-Lagrange equation  $(d/dt)(\partial { L}/\partial \dot{\bm r}_1)=\partial {L}/\partial \bm r_1$  yields the  same dynamics as found in  Eqs.~(\ref{1}) and (\ref{2}).

Note the unusual feature that the equations of vortex dynamics are first order in time, with no term associated with vortex mass and  acceleration.  For a system of many vortices in unbounded space, one can define a Hamiltonian $H = \frac{1}{2} \sum_{i\neq j}\tilde V_{ij}(r_{ij})$ that depends on all the vortex coordinates~\cite{Lin43}.
The equations of vortex dynamics have a Hamiltonian form with $x_i$ and $y_i$ as canonical coordinates.  In the presence of boundaries, the factor $\ln(1/r_{ij})$ is replaced by the appropriate Green's function  $G(\bm r_i,\bm r_j)$ that satisfies the relevant boundary condition~\cite{Fett66a}.

This description is  readily generalized to include a type-II superconductor.  In general,   a superconductor can screen a static magnetic field beyond the characteristic  London penetration length $\lambda_L=c/\Omega_s$, where $\Omega_s = \sqrt{n_s e^2/m_e\epsilon_0}$  is the effective superconducting plasma frequency defined with the superconducting electron density $n_s$~\cite{Fett71}.   A type-II superconductor is one for which the London penetration length $\lambda_L$ is larger than the vortex core radius $\xi$.  In such a superconductor, the magnetic field penetrates the material  as an array of quantized flux lines (charged vortices). The interaction between two flux lines is logarithmic for small separations $r_{ij} \ll \lambda_L$ but it decays exponentially for separations large compared to  $\lambda_L$~\cite{Fett66b}.  Apart from overall factors, the interaction energy is proportional to the Bessel function $K_0(r_{ij}/\lambda_L)$. Since $\lambda_L^2\propto 1/e^2$, where $- |e|$ is the electronic charge, a neutral superfluid can be considered  the limit of a type-II superconductor as $e^2 \to 0$ and $\lambda_L\to \infty$~\cite{Fett66b}.

A similar description also holds for two-dimensional vortices in thin superconducting films, as first discussed by Pearl~\cite{Pear64} and subsequently expounded by de Gennes~\cite{deGe66}.  In this thin-film geometry, the point vortices interact mainly through the fringing magnetic fields in the surrounding vacuum.  Hence   the long-range interaction potential varies like $1/r_{ij}$, intermediate between the $\ln r_{ij}$  dependence of a neutral superfluid and the $\exp(-r_{ij}/\lambda_L)$ dependence of a bulk type-II superconductor~\cite{Fett67}.  This latter paper also  contains a general discussion of the relation between the hydrodynamic view that each vortex moves with the local superfluid velocity and the energy view based on the interaction potential and the Magnus effect.

\section{Domain Wall of Relative Phase}\label{domain}

Son and Stephanov~\cite{Son02} emphasize that  two uniform interacting condensates have two basic normal modes, analogous to those of two coupled pendula, namely in-phase and out-of-phase.  In the first mode, the  total density $n$ couples strongly to the overall phase $S_1+S_2$;  in the second mode,  the density difference $n_1-n_2$    couples strongly to the relative phase $S_1-S_2$.

For the in-phase mode, the Euler-Lagrange equation for the overall phase yields a  conservation equation involving  the density-weighted mean phase gradient  $n_1 \bm\nabla S_1+n_2\bm\nabla S_2$.  Correspondingly, the Euler-Lagrange equation for $n$ yields a Bernoulli-like equation.  Taking plane-wave amplitudes $\propto e^{i(\bm k\cdot\bm r -\omega t)}$ and ignoring the small coupling to the out-of-phase mode   give the expected Bogoliubov dispersion relation
$\hbar^2\omega_k^2 \approx 2\epsilon_kng + \epsilon_k^2,$
where $\epsilon_k = \hbar^2 k^2/(2M)$ and $\delta g_{12}$ is ignored relative to the much larger $g$.  The long-wavelength dispersion relation is linear, with the usual speed of sound $v=\sqrt{ng/M}$, and the crossover between the two terms determines the healing length $\xi =\frac{1}{2}\hbar/\sqrt{Mng}\sim 0.2\ \mu$m quoting the typical value from SS at the end of Sec.~II (note that their definition for $\xi$ is smaller by a factor $\sqrt 2$ than the conventional one given near the end of Sec.~\ref{dyn}).

As emphasized by SS, the out-of-phase mode is more unusual, for it involves the Rabi coupling that depends on $\hbar\Omega\sqrt{n_1n_2}\, \cos(S_1-S_2)$.  A similar procedure for $\Omega = 0$  again gives a Bogoliubov dispersion relation with a smaller squared speed of sound $v_{12}^2 \approx 2(\delta g_{12}/M)\,n_1n_2/n$, involving the quantity $\delta g_{12} = g-g_{12}$ instead of the usual interaction constant $g$.  The corresponding  healing length now becomes $\xi_{12} \approx \hbar\sqrt{n/(8M\delta g_{12}n_1n_2)}\sim 3\ \mu$m, again taking the value from SS.  When the coherent Rabi coupling $\Omega$ is added, the out-of-phase mode acquires a frequency gap $\propto \sqrt{\Omega \,\delta g_{12}n/\hbar}$ for small $\Omega$~\cite{Son02}.

In Sec.~III of SS, they study a model with constant and uniform three-dimensional densities $n_1$ and $n_2$, focusing on the variations in phases over length scales large compared to $\xi_{12}$.  The resulting energy-density functional follows directly from Eq.~(\ref{L2})
\begin{multline}\label{Edom}
{\cal E}[S_1,S_2] = \frac{\hbar^2}{2M}\left[ n_1\left(\bm\nabla S_1\right)^2 +n_2\left(\bm\nabla S_2\right)^2\right]\\
-\hbar\Omega\sqrt{n_1n_2}\cos(S_1-S_2).
\end{multline}

The two phases  $S_1$ and $S_2$ obey coupled sine-Gordon equations that occur, for example, in Josephson's phenomenological field theory  of  the phase difference between two superconducting half spaces separated by a thin insulating layer~\cite{Jose65}.  In particular, a one-dimensional domain wall $S_1-S_2 = S_{12}$ has the simple analytic expression
\begin{equation}\label{sineG}
S_{12}(y) = 4\arctan{e^{ky}}, \quad \hbox{with}\quad k^2 =\frac{M\Omega}{\hbar} \frac{n}{\sqrt{n_1n_2}},
\end{equation}
where $y$ is the coordinate perpendicular to the domain wall.
The thickness $k^{-1}$ of the domain wall is comparable with the Rabi oscillator  length $l_\Omega=\sqrt{\hbar/(M\Omega)}$, which here sets the basic length scale.  If the relative phase starts at 0 for large negative $y$, then the net change in relative phase across the domain wall is $2\pi$.  It is not difficult to show that the domain wall has a  surface tension (energy per unit area)
\begin{equation}\label{sigma}
\sigma = 8\,\hbar\Omega\,l_\Omega \,n\left(\frac{n_1n_2}{n^2}\right)^{3/4}=8\,\sqrt{\frac{\hbar^3\Omega}{M}} \,n\left(\frac{n_1n_2}{n^2}\right)^{3/4} ,
\end{equation}
which is Eq.~(25) of SS.

Toward the end of Sec.~III, SS  point out that their approximation of  uniform densities $n_1$ and $n_2$ fails when $l_\Omega\lesssim \xi_{12}$, since the full energy functional allows the domain wall to unwind.  Their App.~A studies this problem of metastability in detail, confirming the above qualitative estimate.

The coherent  coupling also can  induce time-dependent Rabi oscillations between the two states $\psi_1$ and $\psi_2$, as discussed briefly in SS below their Eq.~(10)  and seen experimentally, for example, in~\cite{Matt98}.  SS include a related effect in their  study of the stationary domain wall  of relative phase (Sec.~IV), where the total  current is conserved, with  the  currents of the two components having opposite contributions that cancel.
Our Sec.~\ref{coherenttrap} studies the corresponding behavior for a single trapped vortex in a two-component coherently coupled condensate. Here, the vorticity transfers coherently and periodically between the two condensates, with no associated population transfer.  Section~\ref{JosephsonDynamics} studies the population and vorticity transfer in more detail.

\section{Two Vortices in Two  Unbounded Coherently Coupled  BECs}\label{2BEC}

How does this Rabi-coupling energy affect the motion of one or more vortices in a uniform two-component BEC?  
In the following, we use the time-dependent variational Lagrangian formalism to provide   approximate  answers  in both  limits of large $l_\Omega=\sqrt{\hbar/(M\Omega)}$ (namely weak Rabi coupling) and small $l_\Omega$ (namely strong Rabi coupling).

For weak coupling,   assume that each component $\psi_j = \sqrt{n_j}e^{iS_j}$ has a vortex with winding number $q_j = \pm 1$ at the two-dimensional position $\bm r_j$, with phase given in Eq.~(\ref{phasej}).
In the absence of coherent Rabi coupling, each vortex has the familiar phase pattern with radial lines of constant phase stretching outward from the source at $\bm r_j$.  The kinetic energy of each vortex appears separately in Eq.~(\ref{L2}), so that they are  uncoupled, apart from the small effect of their well-separated cores.
A weak Rabi coupling with $l_\Omega \gg r_{12}$ changes  this picture only for large distances, distorting the vortex phase patterns to link the two vortices with a domain wall of large thickness $\sim l_\Omega$.  In this limit, use the unperturbed phases to compute the coupling energy [an integral of $\hbar\Omega\sqrt{n_1n_2}\cos(S_1-S_2)$ over the two-dimensional space].  The resulting coupling energy $E_\Omega$ is positive and proportional to $r_{12}^2$ with logarithmic corrections, leading to an attractive force $\bm F\propto \bm r_{21}$.

 In contrast, the strong-coupling  energy $E_\Omega\approx \sigma r_{12}$ follows from the SS analysis quoted  above in Eq.~(\ref{sigma}).   We here study how  vortices in  coherently coupled BECs respond to  such forces.
Section III of SS argues that on scales large compared to $\xi_{12}$, the density of each component can be taken as a spatial constant, so that  the relevant parts of  Eq.~(\ref{L2}) become
\begin{multline}\label{Lsimp}
{\cal L}= -\hbar n_1\dot{S}_1-\hbar n_2\dot{S}_2 -\frac{\hbar^2}{2M}n_1\left(\bm \nabla S_1\right)^2 \\
 -\frac{\hbar^2}{2M}n_2\left(\bm \nabla S_2\right)^2 + \hbar\Omega\sqrt{n_1n_2}\cos(S_1-S_2),
\end{multline}
which here omits any trapping potential.

As a simple and interesting example, consider the case of a single vortex in each component at $\bm r_1$ and $\bm r_2$ with circulations $q_1 = \pm 1$ and $q_2=\pm 1$.  The time-dependent part of the Lagrangian obtained by integrating  (\ref{Lsimp}) is like that in
Eq.~(\ref{Lonecomp})
\begin{equation}\label{TT}
T =  \hbar \pi  \left(q_1n_1\, \hat {\bm z} \times \dot {\bm r}_1 \cdot \bm r_1 + q_2n_2 \,\hat {\bm z} \times \dot {\bm r}_2 \cdot \bm r_2\right),
\end{equation}
but the two vortices now exist in two different components, each with its own number density.  In addition, the integral of the kinetic-energy density [the two  terms proportional to $(\bm \nabla S_j)^2$] yields only the two self-energies, for there is no term involving $\bm \nabla S_1\cdot \bm \nabla S_2$.  Hence these terms have no effect on the dynamical motion.  As a result, two vortices, one in each component, remain stationary  unless they are coherently coupled by  the Rabi energy
\begin{equation}\label{EOmega}
E_\Omega = -\hbar\Omega\,\sqrt{n_1n_2}\int d^2 r\cos(S_1-S_2).
\end{equation}

Independent of the strength of the coupling, this Rabi energy $E_\Omega(r_{12})$ acts like a two-dimensional central potential, assuming that  the system is unbounded and uniform (hence translationally invariant). Equations (\ref{TT}) and (\ref{EOmega}) yield the Lagrangian $L = T-E_\Omega$; it  determines the dynamical equation of motion for each vortex.  Vortex 1 in component 1 obeys
\begin{equation}\label{v1}
2\pi\hbar q_1n_1\dot{\bm r}_1 = \hat{\bm z}\times \bm F_1^\Omega,
\end{equation}
where $\bm F_1^\Omega = -\bm\nabla_1 E_\Omega$.  Similarly,
 \begin{equation}\label{v2}
2\pi\hbar q_2n_2\dot{\bm r}_2 = \hat{\bm z}\times \bm F_2^\Omega,
\end{equation}
where $\bm F_2^\Omega = -\bm\nabla_2E_\Omega= -\bm F_1^\Omega$.

By inspection, the motion conserves the vector quantity $q_1n_1 \bm r_1 + q_2n_2\bm r_2$.  In addition, the relative vector $\bm r_{12} \equiv \bm r_1-\bm r_2$ obeys the dynamical equation
\begin{equation}\label{r12}
\dot{\bm r}_{12} = \frac{q_1n_1 + q_2n_2}{2\pi\hbar q_1q_2 n_1n_2} \,\hat{\bm z} \times \bm F_1^\Omega.
\end{equation}

As a simple example, consider two positive vortices with $q_1=q_2 = 1$. In this case, the corresponding density-weighted centroid $ n_1\bm r_1 + n_2\bm r_2$ remains fixed.  In contrast, the relative vector $\bm r_{12}$ satisfies
\begin{equation}
\dot{\bm r}_{12} = \frac{n}{2\pi\hbar n_1n_2}\hat{\bm z}\times \bm F_1^\Omega,
\end{equation}
but the details depend on the explicit form of the Rabi coupling energy $E_\Omega(r_{12})$.

More generally, for two vortices with unit charges $q_1$ and $q_2$, the  center of motion 
 $\bm r_{\rm cm}  = \frac{1}{2}(\bm r_1 + \bm r_2)$ obeys the dynamical equation 
\begin{equation}\label{vp}
\bm {\dot} {\bm r}_{\rm cm}  = \frac{1}{4\pi \hbar} \,\frac{q_1n_2 - q_2 n_1}{n_1n_2} \hat{\bm z}\times \bm F_1^\Omega.
\end{equation}
Specifically, for a vortex pair/vortex dipole with $q_1=- q_2 = 1$, this result reduces to 
\begin{equation}\label{vp1}
\bm{\dot} {\bm r}_{\rm cm}  = \frac{1}{4\pi\hbar} \,\frac{n}{n_1n_2}\hat{\bm z}\times \bm F_1^\Omega,
\end{equation}
leading to a uniform translation perpendicular to the relative vector $\bm r_{12}$.

\subsection{Weak Rabi coupling}\label{weak}

If the coherent coupling is weak, namely if $l_\Omega=\sqrt{\hbar/(M\Omega)}$ is large compared to the intervortex separation $r_{12}$ (and $r_{12}\gg \xi_{12}$), then the phase pattern of each vortex can be taken as undisturbed over physically relevant distances. Thus,  evaluate the Lagrangian per unit length $L =  \int d^2r {\cal L}$ by integrating over an unbounded two-dimensional Rabi-coupled two-component condensate.

In the present limit of weak Rabi coupling, it suffices to compute the energy $E_\Omega^\pm$ of the coherent coupling using the unperturbed phases of each component, where the product $q_1q_2 = \pm 1$ determines the sign $\pm$.  Thus,  it is necessary to evaluate the integral
\begin{equation}\label{EOm}
E_\Omega^\pm = \hbar\Omega\sqrt{n_1n_2}\int d^2 r \left[C_{\pm}-\cos\left(S_1-S_2\right)\right],
\end{equation}
where $C_{\pm}$ is a constant that eliminates the leading divergence of the integral;  it depends only on the product $q_1q_2$:    $C_+ = 1$ but $C_{-}=0$.

Comparison with Eq.~(\ref{phasej}) shows that
\begin{equation}\label{trig}
\cos S_j = \frac{x-x_j}{|\bm r-\bm r_j|}\quad\hbox{and}\quad \sin S_j = q_j\frac{y-y_j}{|\bm r-\bm r_j|}.
\end{equation}
To simplify the calculation, choose $(x_1,y_1)= (-\frac{1}{2}r_{12},0)$ and $(x_2,y_2)= (\frac{1}{2}r_{12},0)$, so that the two vortices are symmetrically placed on the $x$ axis  with separation $r_{12}$.  As a result,
\begin{equation}\label{cos}
\cos(S_1-S_2) = \frac{x^2-\frac{1}{4}r_{12}^2 \pm y^2}{\sqrt{\left(x^2+\frac{1}{4}r_{12}^2 +y^2\right)^2 -x^2r_{12}^2}}
\end{equation}
 To evaluate $E_\Omega^\pm$ in (\ref{EOm}),  use plane polar coordinates $x=r\cos\theta$ and $y=r\sin\theta$, and introduce the dimensionless variable $u=2r/r_{12}$, so that
\begin{widetext}
\begin{equation}\label{Epm}
E_\Omega^\pm = \frac{\hbar\Omega}{4}\sqrt{n_1n_2}\,r_{12}^2\int_0^{u_0} udu\int_0^{2\pi}d\theta\left[C_\pm -\frac{u^2\left(\cos^2\theta \pm\sin^2\theta\right)-1}{\sqrt{\left(u^2+1\right)^2-4u^2\cos^2\theta}}\right].
\end{equation}
\end{widetext}
Here the radial integral diverges logarithmically at the upper limit and $u_0$ is a  cutoff parameter.

The angular integral can be evaluated in terms of complete elliptic integrals, and use of Landen's  transformation in  the Appendix gives
\begin{equation}\label{E+}
E_\Omega^+\approx \frac{\pi}{2} \,\hbar\Omega\sqrt{n_1n_2}\,r_{12}^2\ln\left(\frac{4\Lambda}{r_{12}}\right),
\end{equation}
for two vortices with the same sign, where $\Lambda$ is a large-distance cutoff, either the size of the container or  the condensate.

 A similar expansion  for two vortices with opposite signs yields
 \begin{equation}\label{E-}
 E_\Omega^- \approx  \frac{\pi}{4}\,\hbar\Omega\sqrt{n_1n_2}\, r_{12}^2 \ln\left(\frac{5.1361\Lambda}{r_{12}}\right).
\end{equation}
Apart from the logarithmic cutoff, the dominant behavior is a  quadratic (harmonic) dependence on  the separation $r_{12}$ of the vortices. Note that both results are positive and attractive (they differ by roughly  a factor of 2).

Let
$
\bm F_1^\Omega = -\bm\nabla_1E_\Omega^\pm$
be the force on vortex 1 arising from the Rabi coupling.  This force acts along $-\bm r_{12}$, toward vortex 2 and is always attractive.     This behavior is quite different from that for two vortices in classical hydrodynamics (or in a one-component  condensate), where the potential in Eq.~(\ref{Vij}) is  proportional to $q_1q_2 \ln(1/r_{12})$, namely positive and repulsive for $q_1q_2 =1$, but negative and attractive for $q_1q_2 = -1$.

To be specific, consider two positive vortices.  The vector 
$\bm r_{12}$  rotates around
$(n_1\bm r_1+ n_2\bm r_2)/n$
in a  {\it negative} (clockwise) sense at an angular velocity
\begin{equation}\label{rot1}
\Omega_{\rm rot}=- \frac{\Omega n}{2\sqrt{n_1n_2}}\,\left[ \ln\left(\frac{4\Lambda}{r_{12}}\right)-1\right]\approx
 -\Omega \ln\left(\frac{4\Lambda}{r_{12}}\right)
\end{equation}
where the last form holds for $n_1=n_2=n/2$, and for $\Lambda/r_{12}\gg 1$.
 This rotation is {\it opposite} to the sense of rotation for  two positive vortices in classical hydrodynamics.
 As we will see below, we and Ref.~\cite{Tylu16} also find a similar behavior in the strong-coupling limit.

Next  consider the two-component analog of a vortex pair with $q_1=1$ and $q_2 = -1$.  In this case, the  density-weighted vector $n_1\bm r_1-n_2\bm r_2$ remains constant.  In addition, Eq.~(\ref{vp1}) shows that 
the center of motion $\bm r_{\rm cm}$  moves  according to $ \dot{\bm r}_{\rm cm} = -\frac{1}{8}\Omega\,(n/\sqrt{n_1n_2})\, \hat{\bm z}\times \bm r_{12}[\ln(5.1361\Lambda/r_{12})-1]$,
 namely in the direction of flow between the two vortices.  This motion has the {\it same} sense  a vortex pair/dipole in classical hydrodynamics, but note that $\bm r_{12}$ itself rotates according to Eq.~(\ref{r12}) unless $n_1=n_2$.

This dynamical motion  for one vortex in each component arises from the effective quadratic dependence on $r_{12}$ in Eqs.~(\ref{E+}) and (\ref{E-}). The present approximation that the phase field of each vortex extends far beyond their separation distance can be considered a variational trial function for the Lagrangian $L$. Hence this behavior should hold for $l_\Omega\gtrsim  r_{12}\gtrsim\xi_{12}$.  As the Rabi frequency increases (and the Rabi oscillator length $l_\Omega$ decreases), however, the situation becomes quite different, because the domain wall of relative phase significantly distorts the separate vortex phase patterns over the scale $l_\Omega = \sqrt{\hbar/(M\Omega)}$.

\subsection{Strong Rabi coupling}

It is interesting also to consider the case of strong Rabi coupling, when $l_\Omega\lesssim r_{12}$.  In this limit, the phase difference $S_1-S_2$ is confined to the domain wall, and the Rabi energy becomes $E_\Omega \approx \sigma r_{12}$, where $\sigma$ is the surface energy in Eq.~(\ref{sigma}).  Correspondingly, the resulting force on vortex 1 is $\bm F_1^\Omega=-\bm \nabla_1 E_\Omega = -\sigma\,\bm r_{12}/r_{12}$, again attractive and along the vector $-\bm r_{12}$.  For two positive vortices, one in each component, Eq.~(\ref{r12}) gives
\begin{equation}\label{large}
\dot{\bm r}_{12} = - \frac{n}{n_1n_2}\frac{\sigma}{2\pi \hbar\, r_{12}} \hat{\bm z}\times \bm r_{12},
\end{equation}
which predicts a rotation rate
\begin{equation}\label{rot2}
\Omega_{\rm rot}= -\frac{\sigma n}{2\pi\hbar n_1n_2 r_{12}}=- \frac{4\sqrt 2}{\pi}\frac{\Omega l_\Omega}{r_{12}},
\end{equation} in the {\it  negative} (clockwise) sense.  Here, the last form holds for $N_1=N_2 = N/2$.

 Note that this result describes a uniform unbounded condensate.  The Trento group~\cite{Tylu16}  studies two such vortices symmetrically placed  in a harmonic trap and finds the  same result for the rotation frequency $\Omega_{\rm rot}$ in the strong-coupling limit [see their Eq.~(5), and note that their $d$ is $\frac{1}{2} r_{12}$].  In this strong-Rabi-coupling limit,  the trap  has negligible effect on the dynamics.  Such agreement lends credence to the present Lagrangian approach.

It is also interesting to consider two oppositely charged  vortices (a vortex pair/vortex dipole, with $q_1=- q_2 = 1$).  In  the presence of coherent coupling, they  move uniformly in the same direction as  classical vortices do  because both situations involve  attractive forces. Specifically, in the strong-coupling regime when the Rabi coupling energy is $E_\Omega(r_{12}) \approx \sigma r_{12}$, Eq.~(\ref{vp1})   readily yields the translational speed of the pair  
\begin{equation}\label{vpair}
v_{\rm pair} = \frac{\sigma}{4\pi\hbar}\frac{ n}{ n_1n_2} = \frac{2\sqrt 2\,\Omega\, l_\Omega}{\pi} = 4\sqrt 2\,\frac{l_\Omega}{T},
\end{equation}
where the last two results hold for $n_1=n_2 =n/2$, and $T= 2\pi/\Omega$ is the Rabi period.

Neely {\it et al.}~\cite{Neel10} have observed similar dynamical motion for a vortex pair/dipole  in  single-component $^{87}$Rb disk-shaped condensate.  In practice, the boundaries tend to dominate the dynamics: in the single-component case, as the pair approaches the TF radius, the vortices separate and follow the boundary, eventually reuniting on the opposite side.  This periodic motion has been observed for one full cycle.


\subsection{Numerical results}
The simulations we show below have been obtained exploiting a Trotter-Suzuki solver we recently developed~\cite{Witt13,Witt15}. The Trotter-Suzuki formula provides an approximation to the operator evolution that preserves its unitarity, while having a low computational complexity. This results in a stable, high precision and fast evolution. The code is publicly available under an open source license and it is written in C++, with a Python wrapper for ease of use~\cite{TSCode}.
The code has been optimized to use parallel and distributed computational resources providing an almost linear scaling across the nodes of a super-computer.
Nonetheless, most of the results presented here are obtained on a standard desktop machine. To facilitate reproduction of the results, a complete computational appendix is available online~\cite{TSNotebook}.

In this section, we wish to study the motion of two vortices in a uniform two-component BEC, one vortex per component. We consider equal populations $N_1=N_2=N/2$, equal masses and equal intra-component interaction constants ($g_{11}=g_{22} \equiv g$), and vanishing inter-component interaction constant ($g_{12}=0$). 
For numerical purposes, we enclose the two components in a circular well with a hard wall located at radius $R$. We chose the radius $R$ to be much greater than the vortex separation $r_{12}$, and we considered relatively strong interactions $g_{ii}$, so that the vortex core radius $\sim\xi$ is smaller than $r_{12}$. In this way, we ensure that the two vortices are well separated from each other, and move  in  a relatively flat density profile.

\begin{figure*}
\centering
\vskip 0 pt
\includegraphics[width=.98\columnwidth,clip]{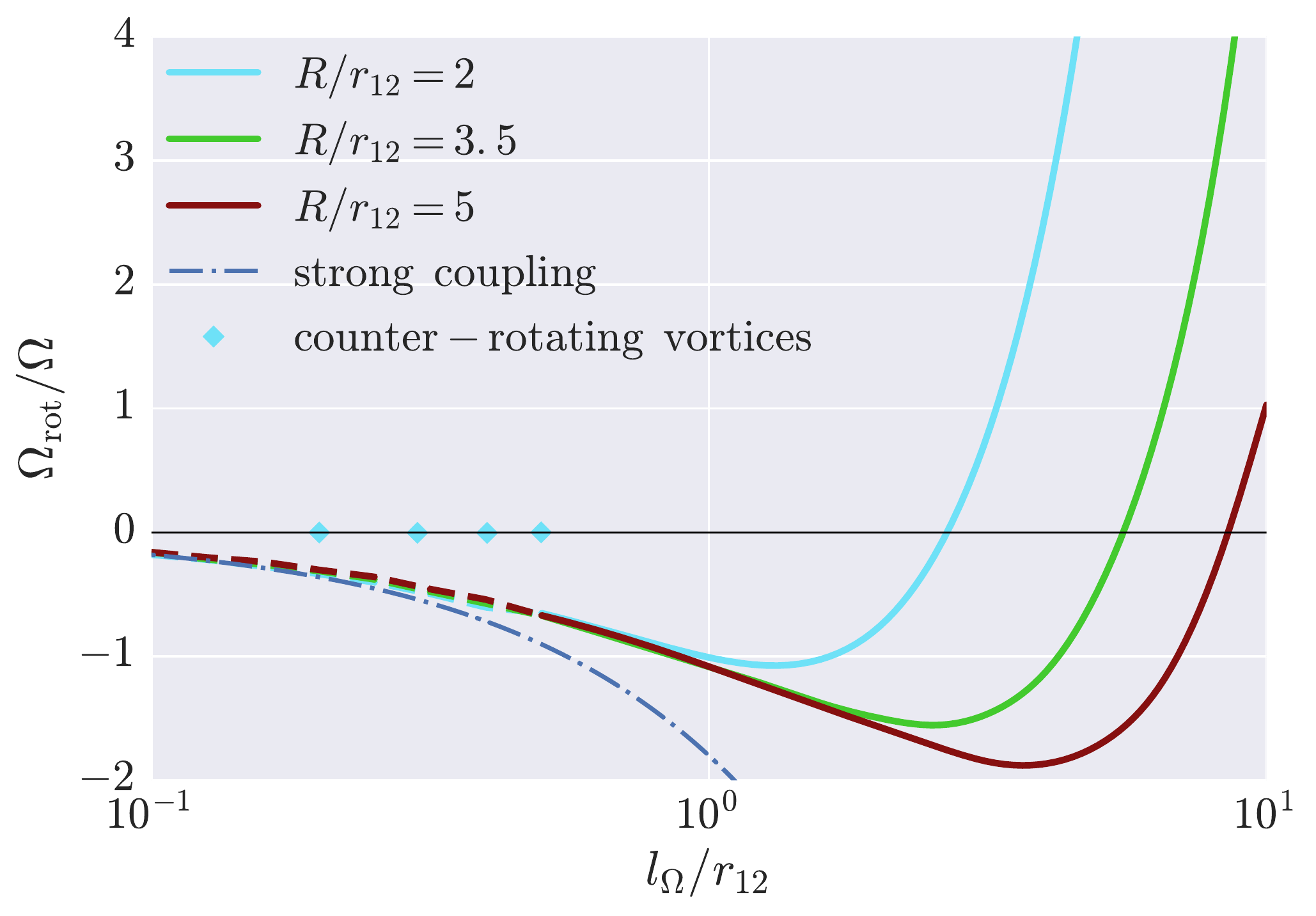}
\includegraphics[width=\columnwidth,clip]{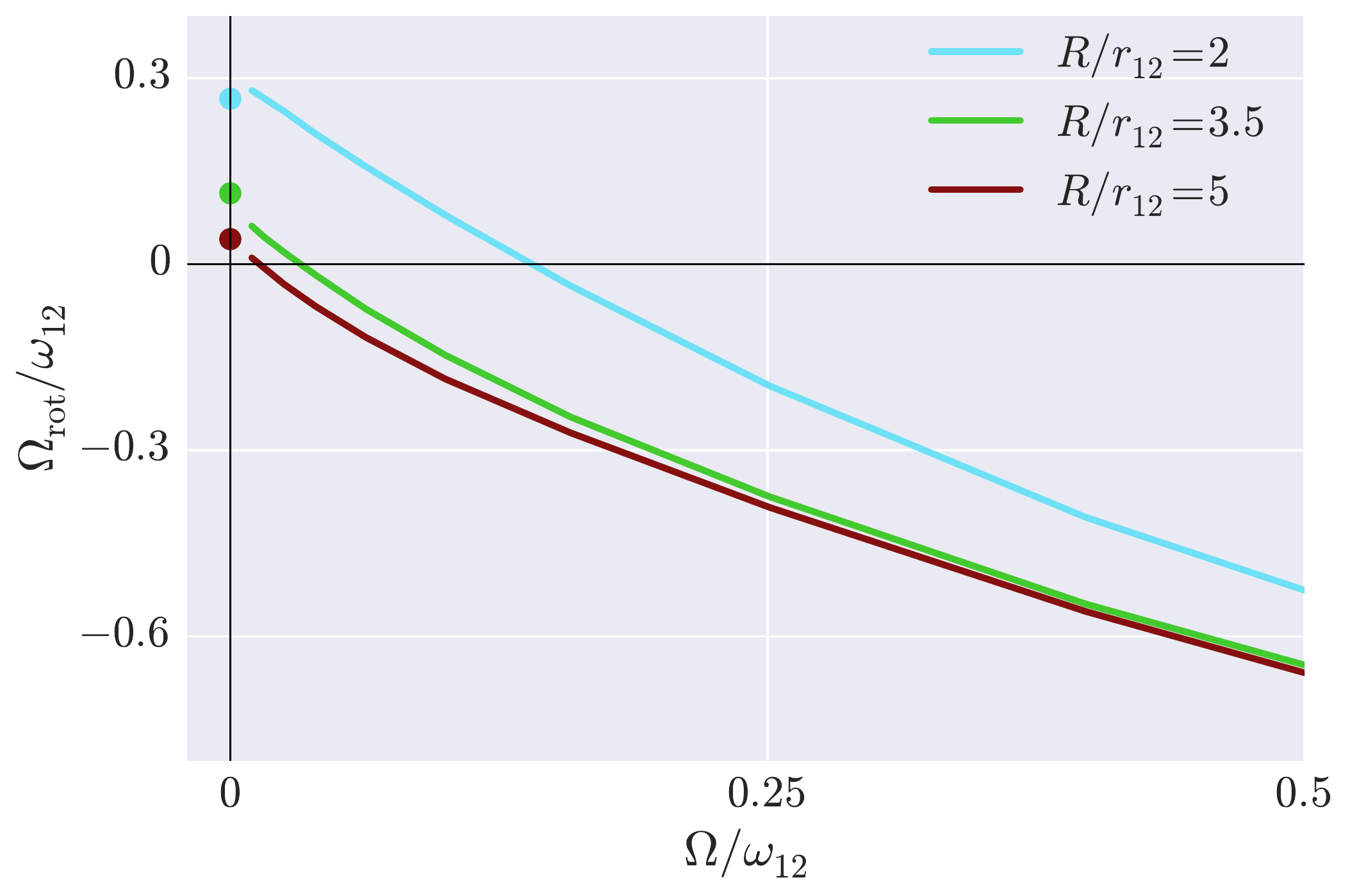}
\caption{{\bf (Left)}  Precession frequency of two co-rotating vortices, one per component. Colors indicate results obtained for different radii $R$ of the circular container; solid lines are results with $\xi=r_{12}/10$, and dashed ones results with $\xi=r_{12}/40$. The dash-dotted line is the strong-coupling limit, Eq.~\eqref{rot2}.
The diamonds are instead results for counter-rotating vortices: for large Rabi coupling vortex-antivortex pairs move uniformly, without precessing. 
{\bf (Right)} The same data are plotted with different axes, to highlight the behavior at weak-coupling.
Here $\omega_{12} \equiv \hbar/M r_{12}^2$, and the dots are the result expected for a single vortex in a single component BEC inside a cylinder, Eq.~\eqref{singleVortexInsideACylinder}.
}
\label{fig:precessionFreqTwoVortices}
\end{figure*}

We initialize the system with two co-rotating vortices located symmetrically across the center of the container, at positions $(\pm r_{12}/2 , 0)$. The vortices, and the corresponding domain wall in the relative phase between them, are obtained performing a short imaginary time evolution, which proceeds along the following steps: (i) we start with normalized wavefunctions which take a constant value inside the circular well, $\psi_1 = \psi_2 =\sqrt{1/\pi R^2}$, and vanish outside; (ii) we phase-imprint two co-rotating vortices, one per component, so that $\psi_j \rightarrow e^{iS_j}\psi_j$, where $S_j$ is given in Eq.~\eqref{phasej}, and $\bm r_1 = -\bm r_2 = \bm r_{12}/2$; (iii) we start the imaginary time evolution, in the presence of an additional pinning potential (two sharply peaked gaussians centered at $\pm \bm r_1$) aimed at keeping the vortex cores stationary (otherwise, they would approach each other during the imaginary time evolution). Once the gas has stabilized, we remove the pinning potential, and we let the system evolve in real time. The precession frequency $\Omega_{\rm rot}$ of the vector $\bm r_{12}$ is obtained by averaging typically over $\sim5$ full revolutions. Our results are summarized in Fig.~\ref{fig:precessionFreqTwoVortices}.

\begin{figure}
\centering
\vskip 0 pt
\includegraphics[width=\columnwidth,clip]{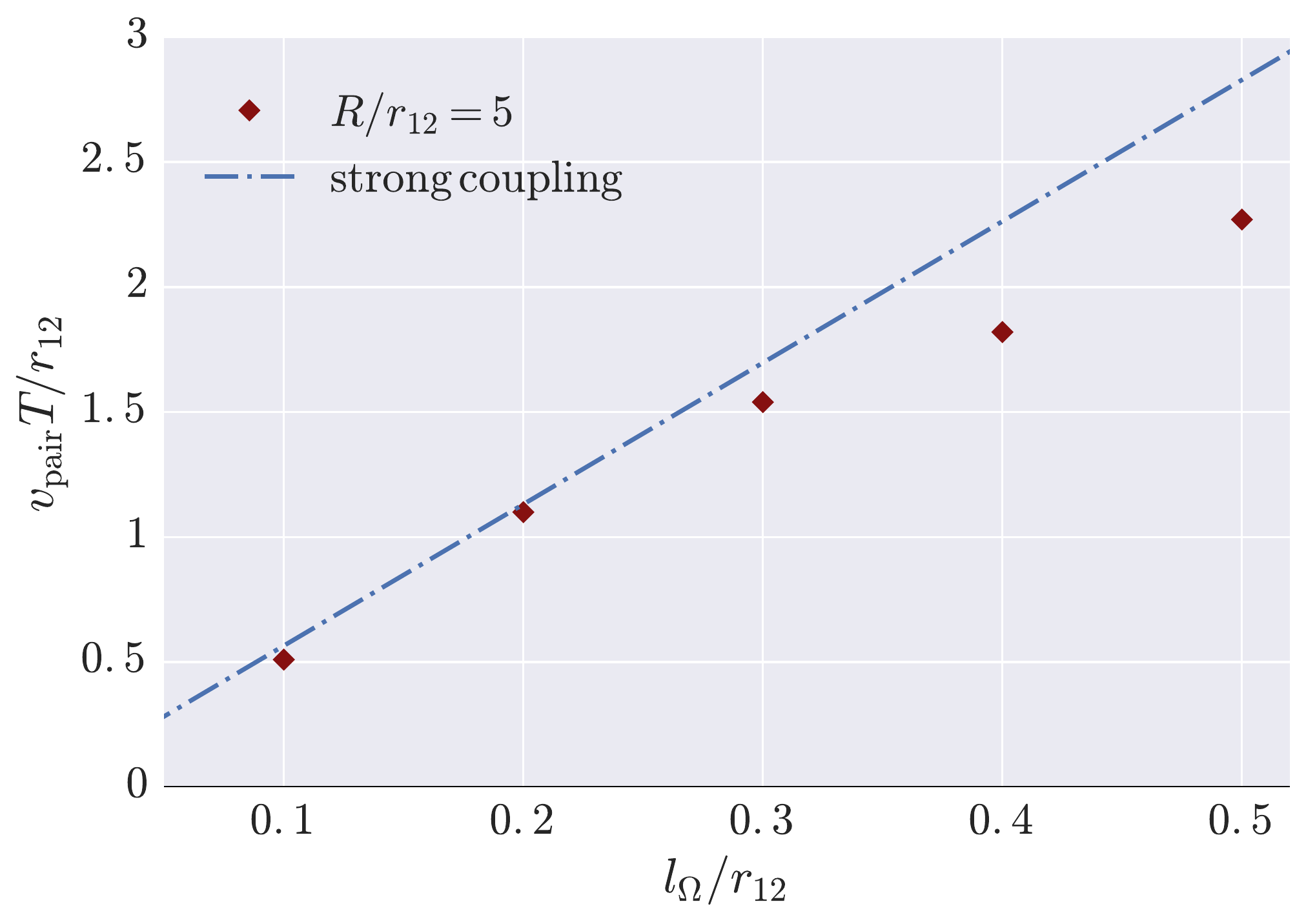}
\caption{\label{fig:translationalVelocityTwoVortices}
Translational velocity of two oppositely charged  vortices, one in each component, in a circular container of radius $R$. The simulation result, shown with diamonds, is compared to the analytical formula Eq.~(\ref{vpair}), valid for strong Rabi coupling.
}
\end{figure}

At strong Rabi coupling, where $l_\Omega\ll r_{12}$, the precession frequency is negative (i.e., the vortices precess in a  direction opposite to the one of their circulation), and it becomes independent of the radius of the container, nicely converging to the analytical prediction, Eq.~\eqref{rot2}. The results are also independent of the strength of the interaction between atoms if  the coherence length is sufficiently small. Indeed, we observe that for very large Rabi frequency ($l_{\Omega}<r_{12}/2$) the domain wall between the vortices rapidly breaks up in the case $\xi=r_{12}/10$, while it remains relatively stable over the whole frequency range studied when  $\xi=r_{12}/40$.
Computations with small $\xi$ are particularly expensive, as they require a very closely-spaced computational grid to sample correctly the rapidly-varying vortex core. In order to always satisfy the inequality $l_{\Omega} \gg \xi$ which ensures stable domain walls, we therefore considered two values of the intra-component interaction constants, depending on the value of $l_{\Omega}$. In particular, we used $\xi=r_{12}/10$ for $l_{\Omega} > r_{12}/2$,  and $\xi=r_{12}/40$ for $l_{\Omega} <  r_{12}/2$.

For  weak coupling, instead, our numerical results indicate a behavior that differs from the one discussed in  Sec.~\ref{weak}.  The presence of a container (necessary in our simulations) rules out the observation of the logarithmic behavior predicted in Eq.~\eqref{rot1}.
At large $l_\Omega$ we find that the precession frequency changes sign and saturates to a small, positive value, somewhat in agreement with that found in~\cite{Tylu16}.
For sufficiently large containers, our data converge to the result expected for a single vortex in a single component BEC, located inside a cylinder, at distance $r_{12}/2$ from its axis. This configuration is discussed, e.g., in Refs.~\cite{Lamb79,Peth08}, and the precession frequency is predicted to be
\begin{equation}\label{singleVortexInsideACylinder}
\Omega_{\rm rot}=\frac{\hbar}{M(R^2-r_{12}^2/4)},
\end{equation}
a result which is displayed with colored dots in the right panel of Fig.~\ref{fig:precessionFreqTwoVortices}.

Finally, we simulated the case of oppositely charged vortices, with $q_1=-q_2=1$. Once more, the simulations reproduce the predicted behavior in considerable detail. In particular, away from the boundaries the vortex dipole evolves with vanishing precession frequency, see diamonds in the left panel of Fig.\ \ref{fig:precessionFreqTwoVortices}. The vortex pair instead translates uniformly, and in the strong-coupling limit its velocity converges to the analytical prediction given in Eq.\ \eqref{vpair}, see Fig.\ \ref{fig:translationalVelocityTwoVortices}. Approaching the edge of the computational grid, where hard wall boundary conditions are imposed, the dynamics gets however more involved. In particular, over some long simulations we observed that two extra vortices are nucleated at the boundary, and enter the condensate. The new vortices, one per component, have charges $q_1=-q_2=-1$, opposite to the charges of the initial vortices. At this point, the relative phase displays two narrow domain walls, one connecting the two vortices with $q_1=q_2=1$, and the other joining the two vortices with $q_1=q_2=-1$. If the two pairs are sufficiently far apart, these will behave independently, each pair precessing smoothly around its own center of mass, as discussed earlier on in this Section. In agreement with theory, the pair of positive (negative) vortices is found to precess in the clockwise (anticlockwise) direction. A video of the complete simulation is available in the Supplemental Material \cite{SuppMat}. Note that this behavior agrees with that predicted by Son and Stephanov~\cite{Son02}, namely that domain walls naturally run between two same-sign vortices, one in each component.

\section{Single Vortex in Trapped Two-component Condensate with Coherent Coupling}\label{coherenttrap}

One other case for coherent  Rabi coupling also merits careful study:  a single vortex at $\bm r_1$ with $|q_1|=1$  in  component one of a trapped Thomas-Fermi two-component condensate.  The nonuniform density arising from the harmonic trap potential  exerts a force on the vortex so that it  precesses in the same sense as its circulation, but the effect of the Rabi-induced harmonic coupling  requires a detailed analysis. In addition, we briefly consider the similar but simpler case of a vortex in one component of a two-component condensate with weak interaction constants, where a Gaussian variational trial function is appropriate.

\subsection{Analytical results for strong-coupling Thomas-Fermi limit}\label{TFvortex}

The normalized two-component trial function here has the form used  in studying the motion of a vortex in a trapped two-dimensional spin-orbit coupled condensate~\cite{Fett14}
\begin{equation}\label{Psirabi}
\Psi(\bm r)= \left(\frac{2}{\pi R^2}\right)^{1/2}\left(1-\frac{r^2}{R^2}\right)^{1/2}
\left(\begin{matrix} \sqrt{N_1}\,e^{iS_1}\\[.2cm]
\sqrt{N_2}\,e^{i\bar\phi} \end{matrix}
\right),
\end{equation}
where $S_1$ is the phase given in Eq.~(\ref{phasej}) for a vortex in component one at position $\bm r_1$ with circulation $q_1$ and $\bar\phi$ is an additional phase, initially taken as constant.  Our numerical studies show clearly, however, that $\bar\phi$ varies linearly with time, and we henceforth assume $\bar\phi = \kappa t$, where $\kappa$ is constant.

The evaluation of the trap energy and interaction energy  for $g_1=g_2=g_{12} = 4\pi a \hbar^2/(\sqrt{2\pi}M d_z)$  is given in~\cite{Fett14},  yielding the variational estimate
\begin{equation}\label{R}
R^4 = \frac{16}{\sqrt{2\pi}}\frac{Nad_\perp^4}{d_z}
\end{equation}
for the Thomas-Fermi condensate radius $R$.  Since these contributions  have no effect on the vortex motion, they are ignored in the subsequent study.

The resulting Lagrangian density (\ref{Lsimp}) now contains only contributions from the time and space varying phase $S_1$, and  the time-dependent term in the Lagrangian follows from Eq.~(11) of Ref.~\cite{Fett14}
\begin{eqnarray}\label{T1}
\nonumber T &=& -\frac{2\hbar N_1}{R^2}\, q_1\left(1-\frac{1}{2}u_1^2\right)\dot{\bm r}_1\times \hat{\bm z}\cdot \bm r_1\\
&=&-2\hbar N_1q_1\dot{\phi}_1\left(u_1^2-\frac{1}{2}u_1^4\right),
\end{eqnarray}
where we omit a constant term arising from the time dependence of $\bar\phi = \kappa t$.
Here  the second form uses plane polar coordinates $\bm r_1=(r_1,\phi_1)$, with  $u_1 = r_1/R$.  Similarly, the kinetic energy of the circulating flow around the vortex follows from Eq.~(10)  of~\cite{Fett14}
\begin{equation}\label{Ekin1}
E_k = \frac{\hbar^2 N_1}{M R^2}\left\{(1-u_1^2)\ln\left[\frac{(1-u_1^2)R^2}{\xi^2}\right] + 2u_1^2-1\right\}.
\end{equation}

\begin{figure*}[ht]
  \begin{center}
  \includegraphics[width=1.5\columnwidth,clip]{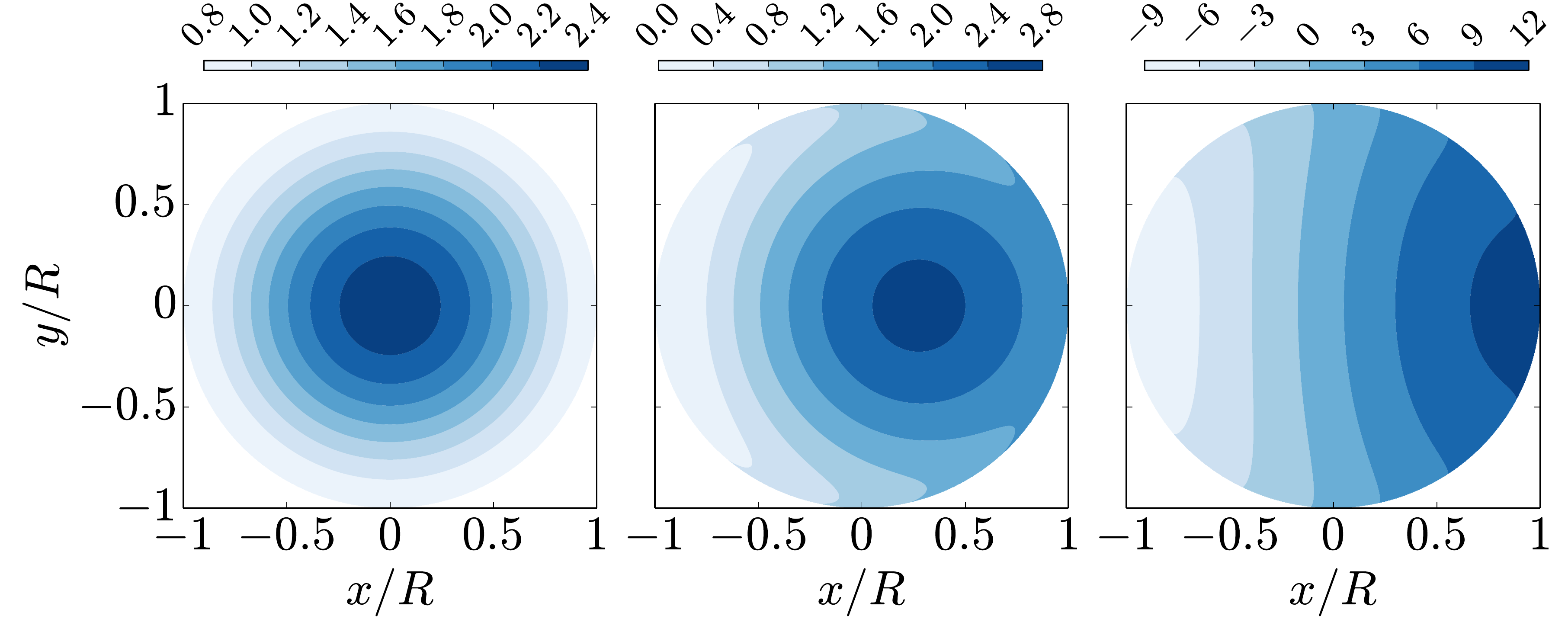} 
  \caption{Contours of equal dimensionless energy $\tilde E$, for $N_1 = N_2$, $\bar\phi=0$, and $R/\xi = 5$.
  From left to right, $M\Omega R^2/\hbar = 0,1,10$. For $\bar\phi\neq0$, the contours would be rotated in the plane by an angle $\bar\phi$.
  }
\label{fig:Econtour}
  \end{center}
\end{figure*}

Finally, the Rabi coupling energy involves the integral of $-\hbar \Omega\sqrt{n_1n_2}\,\cos(S_1-\kappa t)$.  This integral also appears in the study of vortices in  spin-orbit coupled  BECs;  it  is evaluated in Eqs.~(28) and (29) of~\cite{Fett14} for the present case of a half-quantum vortex, namely one vortex in one component and no vortex in the other.  The resulting Rabi coupling energy becomes
\begin{equation}\label{ERabi1}
E_\Omega = \hbar \Omega \sqrt{N_1N_2}\,|f(u_1)|\cos(\phi_1-\kappa t),
\end{equation}
where we have $|f(u_1)| \approx \frac{4}{3}u_1 - cu_1^3$ and $c= 4/3 -128/(45\pi) \approx 0.428$.  Note that $|f(u_1)|$  vanishes for $u_1=0$ and  is positive for $0\le u_1\le 1$ (the relevant range).  Here  the two components have a relative phase $\bar\phi=\kappa t$, and the last factor  $\cos(\phi_1-\kappa t)$ rotates the contours in the center and right part of Fig.~\ref{fig:Econtour} through an angle $ \kappa t$.
 The total energy is the sum $E= E_k + E_\Omega$.  Both terms are positive, but $E_k$ is isotropic and decreases with increasing $u_1$, whereas $E_\Omega$ contains a factor $\cos(\phi_1 -\kappa t)$.

It is convenient to normalize all these terms by the characteristic energy $\hbar^2 N_1/(MR^2)$, in which case we find the dimensionless quantity
\begin{equation}\label{tildeT1}
\tilde{T} = -\frac{2MR^2 }{\hbar }\,q_1\dot{\phi}_1\left(u_1^2-\frac{1}{2}u_1^4\right).
\end{equation}
Similarly, the dimensionless total energy is
\begin{multline}\label{tildeE1}
\tilde{E}(u_1,\phi_1) = (1-u_1^2)\left[2\ln\left(\frac{R}{\xi}\right)+ \ln(1-u_1^2)\right] + 2u_1^2-1\\
+ \frac{M\Omega R^2}{\hbar}\,\sqrt{\frac{N_2}{N_1}} \,|f(u_1)|\cos(\phi_1- \kappa t).
\end{multline}
The dimensionless Lagrangian thus becomes $\tilde L = \tilde T -\tilde E$.

Since $\tilde L$ does not involve $\dot{u}_1$, the Euler-Lagrange equation for $u_1$ takes the simple form $\partial \tilde L/\partial u_1= 0$, which yields the effective precession rate
\begin{eqnarray}\label{dotphi1}
\dot{\phi}_1& =&   -\frac{\hbar }{2MR^2}\,\frac{q_1}{2u_1(1-u_1^2)\,}\frac{\partial\tilde E}{\partial u_1}\notag \\[.2cm] 
\nonumber & = & \frac{\hbar }{MR^2}\,\frac{q_1}{1-u^2_1}\left[\ln\left(\frac{R}{\xi}\right) + \frac{1}{2}\ln(1-u_1^2) -\frac{1}{2}\right] \\
& &-\frac{\Omega q_1}{4u_1(1-u_1^2)} \sqrt{\frac{N_2}{N_1}}\,\cos(\phi_1-\kappa t) \,\frac{d|f(u_1)|}{du_1}.
\end{eqnarray}

The corresponding Euler-Lagrange equation for $\phi_1$ becomes
\begin{eqnarray}\label{dotu1}
\dot{u}_1 &=&  \frac{\hbar }{2MR^2}\,\frac{q_1}{2u_1(1-u_1^2)\,}\frac{\partial\tilde E}{\partial \phi_1}\notag\\[.2cm]
&=& -\frac{\Omega q_1}{4u_1(1-u_1^2)} \sqrt{\frac{N_2}{N_1}}\,|f(u_1)|\,\sin(\phi_1-\kappa t).
\end{eqnarray}

Note that 
$d\tilde E/dt =\left(\partial_{u_1}\tilde E\right)\dot{u}_1+ \left(\partial_{\phi_1} \tilde E\right)\dot{\phi}_1 +\partial_t\tilde E$
 no longer vanishes because of the last term arising from the explicit time-dependence of $\bar\phi$.  Nevertheless, it is still instructive to exhibit  contours of constant $\tilde E$.  Figure~\ref{fig:Econtour}   shows contour plots of $\tilde E$ for $N_1=N_2$, illustrating how the inclusion of the Rabi coupling term affects these energy contours.
Left side is for $M\Omega R^2/\hbar= 0$, with concentric axisymmetric contours; center is for $M\Omega R^2/\hbar=1$, showing displaced nearly circular contours;  and right side is for $M\Omega R^2/\hbar = 10$.  Note that in these latter cases,  some or all trajectories will leave the condensate.
 For small Rabi coupling with $M\Omega R^2/\hbar = R^2/l_\Omega^2\ll 1$, the perturbation has a time-dependent dipolar form $\propto \cos(\phi_1 -\kappa t)$, corresponding to a lateral displacement of the circular contours to first order in the small parameter.

The Amherst group~\cite{Frei10} has developed a valuable thermal-quench technique that creates one vortex in a single-component BEC with probability about 25\%.  There is no obvious reason why this rapid thermal-quench technique should not work for a two-component coherently coupled condensate.  It may be simplest  first to create a half-quantum vortex and then turn on the Rabi coupling, but other experimental options could be preferable.

\subsection{Analytical results for weak-coupling}\label{gaussianvortex}

The previous Sec.~\ref{TFvortex} used the Thomas-Fermi model, which applies to a strong-coupling limit with $R/d_\perp\gg1$, where as before $R$ is the Thomas-Fermi radius and $d_\perp$ is the two-dimensional oscillator length.  The present weak-interaction case involves a quite different approximation,   using the low-lying states of the two-dimensional harmonic oscillator as a basis~\cite{Butt99,Linn99}.

We first examine two  vortex-free components that set the basic energy $E_{\rm GP0}$.  As before, the condensates are assumed to be tightly confined in the axial (perpendicular) direction. 
The axial kinetic and trap energy are an overall shift and will be ignored.  
The energy functional is given by
\begin{multline}\label{EGP}
E_{\rm GP} =\int d^2r \Big[ \frac{\hbar^2}{2M} \left|\bm\nabla_\perp \Psi\right |^2 + \frac{1}{2} M\omega_\perp^2 r^2 |\Psi|^2\\
 + \frac{1}{2}\sum_{ij} g_{ij} n_in_j - \frac{1}{2}\hbar\Omega \left(\psi_1^*\psi_2 + \psi_2^*\psi_1\right)\Big],
\end{multline}
where $\Psi $ is a two component vector with elements $(\psi_1,\psi_2)$ and $n_i = |\psi_i|^2$ is the particle density of the $i$th  component.

In the absence of a vortex,  take a normalized gaussian trial function with a variable radius scaled by the parameter~$\beta$
\begin{equation}\label{Psi0}
\Psi_0(\bm r) =\frac{1}{d_\perp\beta \sqrt\pi}\,\exp\left(-\frac{r^2}{2d_\perp^2 \beta^2}\right) \left(\begin{matrix}  \sqrt{ N_1}\\
\sqrt{ N_2}\end{matrix}\right).
\end{equation}
  Evaluating the ground-state energy is straightforward and gives
\begin{equation}\label{EGP0}
E_{\rm GP0} = \frac{\hbar \omega_\perp N}{2}\left( \beta^2 +\frac{1+{\cal G}}{\beta^2}\right) -\hbar\Omega \sqrt{N_1N_2},
\end{equation}
where the first term (in parenthesis) is the trap energy, the kinetic energy, and the interaction energy,  and the Rabi energy is simply another  constant shift.  The dimensionless  interaction contribution is 
\begin{equation}\label{calI}
{\cal G} =\frac{1}{2\pi d_\perp^2 \hbar\omega_\perp N}\left( g_{11}N_1^2+  g_{22}N_2^2 +2  g_{12}N_1N_2\right).
\end{equation}

Minimization with respect to $\beta$ readily yields the expansion parameter
\begin{equation}\label{beta}
\beta^4 = 1+{\cal G},
\end{equation}
which replaces Eq.~(\ref{R}) for the ratio $R^4/d_\perp^4$ in the TF version. As a variational treatment, this value of $\beta$  is chosen as fixed even in the presence of a vortex.   Note that positive interactions indeed expand the condensate.  In the limit of large ${\cal G}$, the kinetic energy is negligible, and this model becomes a gaussian approximation for the TF limit.

 For a TF condensate, the vortex core size ($\sim \xi\ll d_\perp$) is the small healing length. Hence  the main effect is the phase $S_1$ associated with a vortex.
For the weak-coupling case,  however, the core size is comparable with the trap oscillator length $d_\perp$, which effectively replaces the healing length when $gn\lesssim \mu \sim \hbar \omega_\perp$ in  a one-component condensate.  Use the normalized one-component ground state $\chi_0(r) = (d_\perp\beta\sqrt\pi)^{-1}\exp(-r^2/2d_\perp^2 \beta^2)$, and the first excited state with a central positive vortex  $\chi_{1} (\bm r) = (z/d_\perp\beta) \chi_0(r)$, where $z = x+i y= re^{i\phi}$.  A  linear combination of 
these two states $\propto (z-z_1)\chi_0$ characterizes a single vortex located at $z_1 = x_1+iy_1= r_1e^{i\phi_1}$ in plane-polar coordinates.  Note that this state has a node at $z=z_1$, and the phase of the wave function increases by $2\pi$ on once encircling the node in the positive sense.  

Introduce dimensionless units with $d_\perp$ as the length scale, $\hbar\omega_\perp$ as the energy scale and $\omega_\perp^{-1}$ as the time scale.  In this way, the trial state $\Psi_1(\bm r)$ for a vortex in component 1 located at complex coordinate $z_1$ has the two normalized components 
\begin{eqnarray}\label{Psi1}
\nonumber \psi_1(\bm r) &=& \frac{\sqrt{N_1}}{\beta\sqrt{\beta^2 + r_1^2}\sqrt\pi} \,(z-z_1) \,e^{-r^2/2\beta^2}\\
 \psi_2(r) &=& \frac{\sqrt{N_2}}{\beta\,\sqrt\pi} \,e^{-r^2/2\beta^2}\,e^{i\bar\phi},
\end{eqnarray}
where again $\bar\phi = \kappa t$ based on our numerical studies.

With the same dimensionless variables, and omitting a constant term arising from the time dependence of $\bar\phi = \kappa t$, the time-dependent part of the Lagrangian involves only $\partial \psi_1/\partial t$,  and one finds
 \begin{equation}\label{TWeak}
T = -\frac{N_1\,r_1^2}{\beta^2 + r_1^2}\,\frac{\partial \phi_1}{\partial t};
\end{equation}
this expression should be compared with Eq.~(\ref{T1}) for the TF limit, especially the last form (note that $u_1$ there is effectively $r_1$ here, and that the extra quartic term there reflects the TF condensate profile).

The relevant GP vortex energy for a weak-coupling system is the difference  between the GP energy  $E_{\rm GP1} $ evaluated  with $\Psi_1$ in (\ref{EGP}) and the ground-state energy Eq.~(\ref{EGP0}).
A straightforward analysis gives 
\begin{multline}\label{EGPweak}
E_{\rm GP}^v = \underbrace{\frac{N_1(1+\beta^4)}{2\left(\beta^2+r_1^2\right)}}_{\rm kinetic+trap}-\underbrace{\frac{g_{12}N_1N_2 }{2\left(\beta^2+r_1^2\right) 2\pi} }_{\rm interspecies} -\underbrace{\frac{g_{1}N_1^2 \beta^2}{4\left(\beta^2+r_1^2\right)^ 22\pi} }_{\rm intraspecies}\\
+\underbrace{\Omega\sqrt{\frac{N_1N_2}{\beta^2 + r_1^2}}\,r_1 \cos(\phi_1-\kappa t)}_{\rm Rabi}.
\end{multline}
Thus the total Lagrangian for the vortex dynamics has the same form as in the TF limit
\begin{equation}\label{L}
L = T- E_{\rm GP}^v,
\end{equation}
where $T$ follows from Eq.~(\ref{TWeak}).

The Euler-Lagrange equations for $L$ readily provide the dynamical equations for the motion of the vortex in this weak-coupling model
\begin{eqnarray}\label{dotphiweak}
\frac{d \phi_1}{d t} &=& -\frac{(\beta^2 + r_1^2)^2}{2N_1\beta^2 r_1}\,\frac{\partial E_{\rm GP}^v}{\partial r_1}
\notag\\[.2cm]
\nonumber &=& \frac{1+\beta^4}{2\beta^2} -\frac{g_{12}N_2}{4\pi \beta^2} -\frac{g_1N_1}{4\pi(\beta^2+ r_1^2)}\\[.2cm]
& &-\frac{\Omega}{2}\sqrt{\frac{N_2}{N_1}}\frac{\sqrt{\beta^2+r_1^2}}{r_1}\,\cos(\phi_1-\kappa t)
\end{eqnarray}
and
\begin{eqnarray}\label{dotrweak}
\frac{dr_1}{d t} &=&  \frac{(\beta^2 + r_1^2)^2}{2N_1\beta^2 r_1}\,\frac{\partial E_{\rm GP}^v}{\partial \phi_1} \notag\\[.2cm]
&=&-\frac{\Omega}{2\beta^2} \sqrt{\frac{N_2}{N_1}}\,(\beta^2+ r_1^2)^{3/2} \,\sin(\phi_1-\kappa t).
\end{eqnarray}
If there is no Rabi coupling, then Eq.~(\ref{dotphiweak}) shows that the vortex orbits are concentric circles.  Evidently,  repulsive interactions act to reduce the precession frequency, and the details depend on the assumed values for the set ${g}_{ij}$.  Note that the precession frequency in the TF limit is of order $\hbar/(MR^2)\ln(R/\xi)$, which is much smaller than $\hbar/(Md_\perp^2) = \omega_\perp$, so that such a reduction is to be expected.
It is intriguing to observe that attractive interactions (with negative ${g}_{ij}$) would act to increase the precession frequency.  Whether  such an effect would be observable remains an open question.

\subsection{Numerical results}
To illustrate the coherent oscillations of vorticity induced by the Rabi coupling $\Omega$, we consider a two-component BEC with equal populations ($N_1 = N_2=N/2$), interaction strengths characterized by $g_{12} = g$, in a two-dimensional harmonic trap of frequency $\omega_\perp$. 
We prepare the system by phase-imprinting a single vortex with positive circulation in the center of the first component, and we find the corresponding ground state by performing a short evolution in imaginary time in absence of Rabi coupling, which allows the formation of a vortex core with the suitable profile at the center of the first component. After equilibration, we switch on the Rabi coupling, and let the system evolve in real time for a variable time $t>0$.

\begin{figure*}[p]
\centering
\includegraphics[width=.8\linewidth,clip]{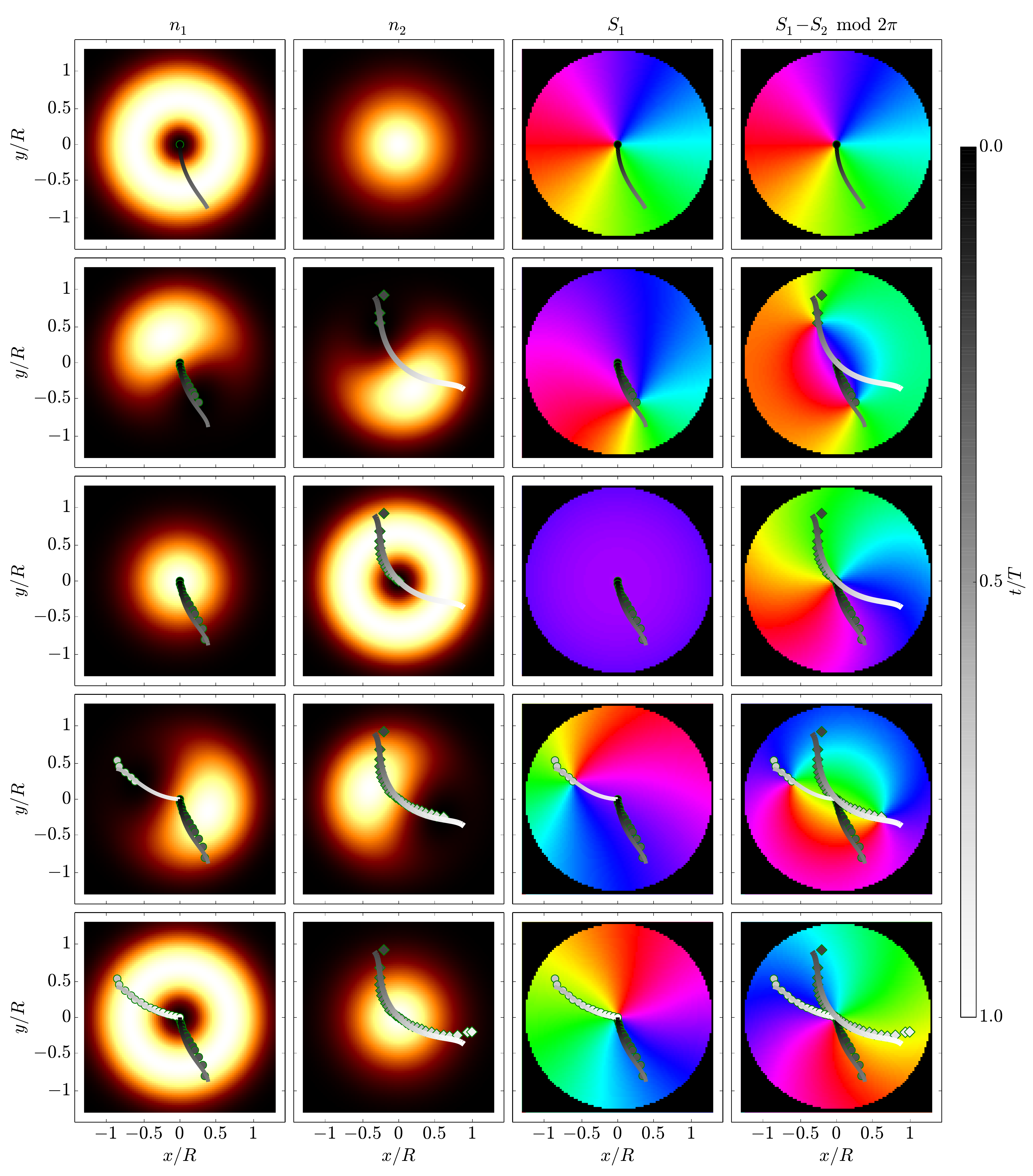}
\caption{Transfer of vorticity in a harmonically trapped two-component BEC, in the  presence of a Rabi frequency $\Omega = 2\omega_\perp$. In this simulation we used a relatively strong coupling, $gN=g_{12}N=40\hbar^2/M$, so that
$R^2/l_\Omega^2 \approx10$ and $(R/d_\perp)^4 \approx 50$. 
From left to right, the columns show, respectively, the particle densities $n_1$ and $n_2$, the phase $S_1$ of $\psi_1$, and the phase difference $S_1-S_2$ at given times: from top to bottom, $t/T=\{0,\,0.25,\,0.5, \, 0.75,\,1\}$, with $T=2\pi/\Omega$ the Rabi period.
Markers indicate the trajectory of the vortex core, up to the time at which the screenshot is taken (circles and squares for vortex core in first and second component respectively). The color of the markers indicate at what (past) time the core was at that specific position.
{\bf Top row:} initial condition, with a vortex at the center of the first component.
{\bf Second row:} the system imaged after one quarter of a Rabi period ($t=T/4$). The vortex core follows the trajectory marked by the symbols: from black ($t=0$) to gray ($t=T/4$), it travels until the edge of the first component; just before $t=T/4$, while the first vortex is still inside the first cloud, a second vortex enters the second cloud, and a domain wall in the relative phase is clearly visible in the fourth column.
{\bf Third row:} after half a Rabi period ($t=T/2$), the first vortex has completely disappeared from the first component, while the new one gradually migrates to the center of the second component.
{\bf Fourth row:} at $t=3T/4$, two vortices are again visible, one in each component, with a domain wall in-between them.
{\bf Bottom row:} after a full Rabi period ($t=T$), the vortex leaves the second component, reappears inside the first, and returns back to its center.
Continuous lines show the predicted trajectory given by Eqs.\ \eqref{dotphi1} and \eqref{dotu1}, for $\kappa = 0.485\omega_\perp$.
A video of the complete simulation, running over five full Rabi cycles, is available in the Supplemental Material \cite{SuppMat}.}
\label{fig:single_vortex}
\end{figure*}

 The dynamics we observe is summarized in Fig.\ \ref{fig:single_vortex}. In the simulation, we have chosen a Rabi coupling such that $R^2/l_\Omega^2 \approx 10$.
At $t=0$, the vortex core starts at the center of the first cloud.
 As time progresses, we observe that the vortex core slowly drifts towards the edge of the first component, and exits the first component to reappear almost simultaneously in the second.
A pair of vortices, one in each component, is actually visible for a brief interval of time centered around $(2n+1)T/4$, with $n=0,1,2,\ldots$, where $T = 2\pi/\Omega$.  After half a Rabi period (i.e., at $t=T/2$, as shown in the central row of the Fig.\ \ref{fig:single_vortex}), the vortex core sits right in the middle of $n_2$, and after a full Rabi cycle the vortex has returned to its starting position, at the center of $n_1$ (bottom row of the Fig.\ \ref{fig:single_vortex}). This coherent transfer of vorticity repeats itself rather uniformly over time. In various simulations, we have for example observed ten complete cycles. Related effects have been discussed theoretically in toroidal traps in Ref.\ \cite{Gall15}, and in harmonic traps in Ref.\ \cite{Tylu16}.

In each panel of Fig.\ \ref{fig:single_vortex}, we plot also the analytical prediction for the trajectory of the vortex (continuous lines), as given by Eqs.\ \eqref{dotphi1} and \eqref{dotu1}.
In this case, the analytical equations were solved using $\kappa = 0.485\omega_\perp$, the value which minimizes the mismatch between the simulated and analytical trajectories over five Rabi periods. 
The vortex core follows very closely the analytical trajectory, 
with a slight mismatch only visible at the border of the condensate, where the Thomas-Fermi approximation is not appropriate.

\begin{figure*}
\centering
\vskip 0 pt
\includegraphics[width=2\columnwidth,clip]{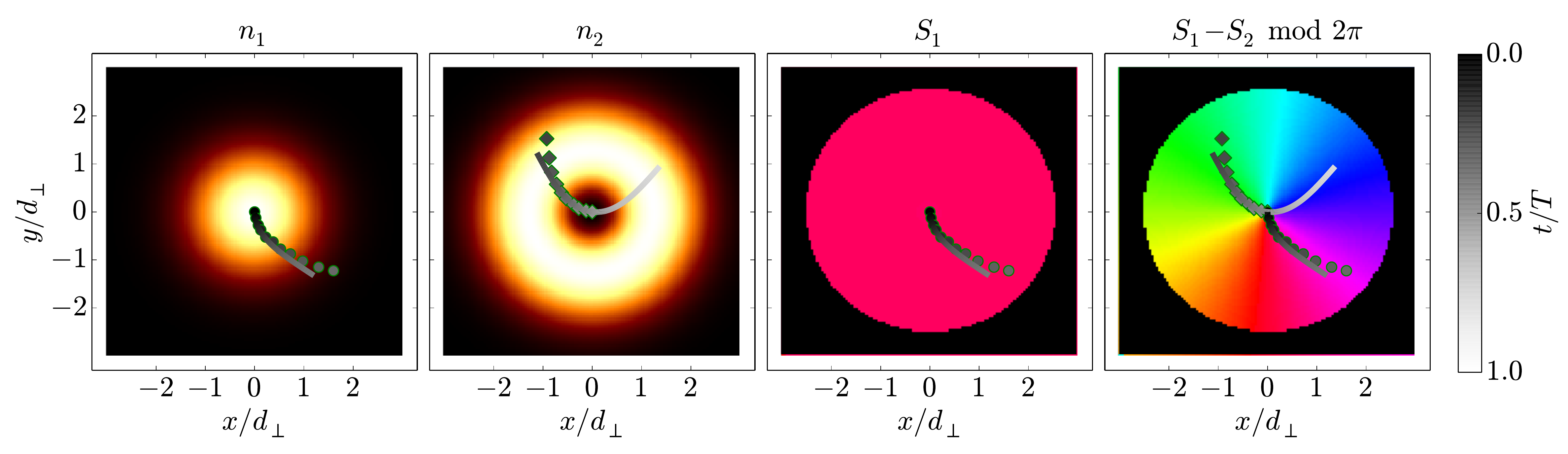}
\caption{Dynamics at weak-coupling ($gN = g_{12}N = 2\hbar^2/M$): from left to right, we show the  particle densities and phases of first and second component of the BEC, after an evolution of half a Rabi period. Markers depict the numerical vortex trajectory, while continuous lines show the predicted trajectory as given by Eqs.\ \eqref{dotphiweak} and \eqref{dotrweak}. The Rabi frequency is set to $\Omega = 2\omega_\perp$, as in Fig.\ \ref{fig:single_vortex}.}
\label{fig:weakCouplingSnapshot}
\end{figure*}

In the case of weaker interactions, we observe a very similar dynamics, displaying coherent transfer of vorticity over many periods. An example of weak coupling dynamics is shown in Fig.\ \ref{fig:weakCouplingSnapshot}.
Here, the analytical trajectories of the vortices are given by the solution of Eqs.\ \eqref{dotphiweak} and \eqref{dotrweak}, and 
we have chosen $\kappa = 0.93$ to minimize the difference between the predicted and simulated trajectories over five Rabi periods.

\begin{figure}
\centering
\vskip 0 pt
\includegraphics[width=\columnwidth,clip]{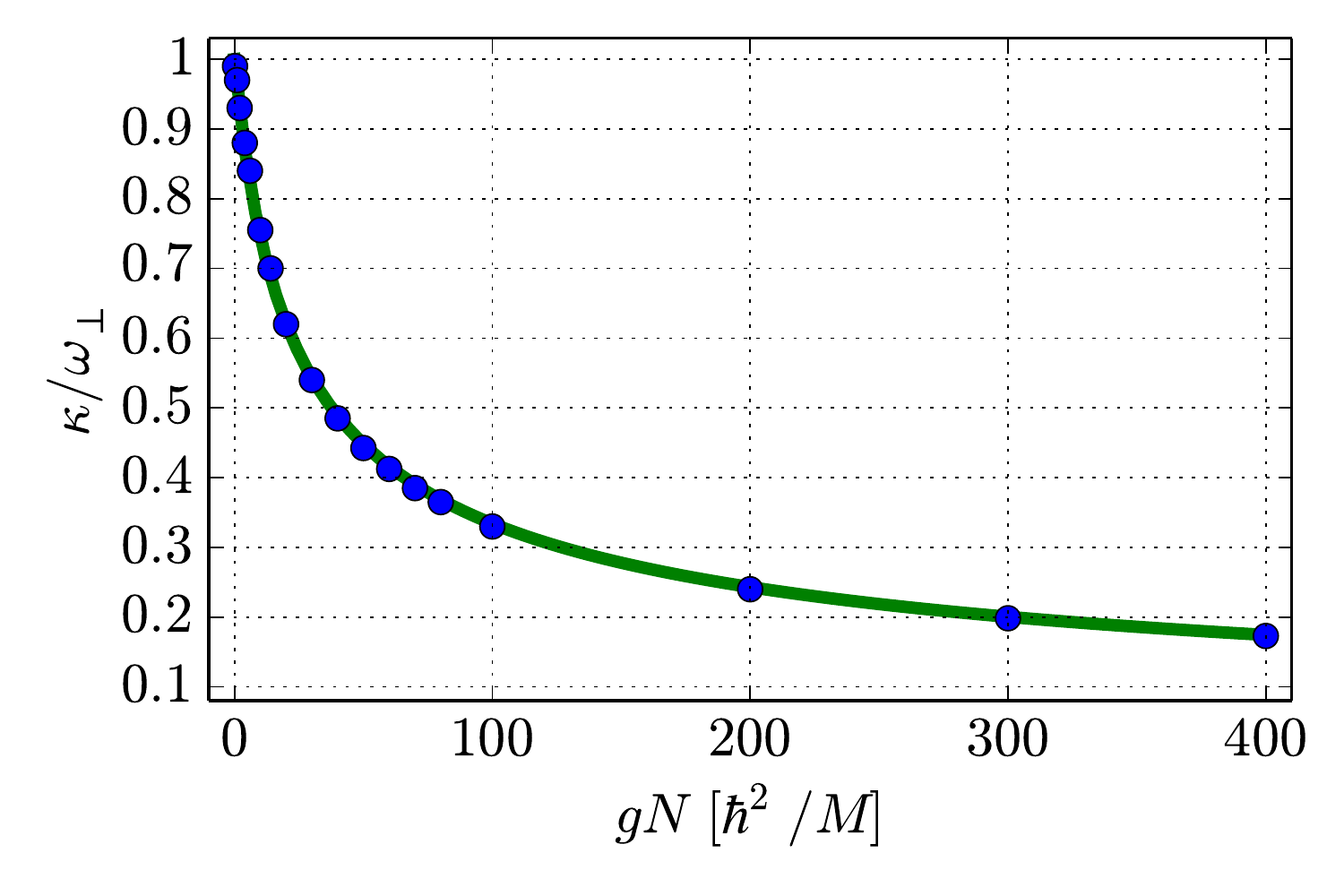}
\caption{Dependence of $\kappa$ on the coupling $gN$. Dots represent the values of $\kappa$ that minimize the difference between the simulated trajectories, and the ones predicted by Eqs.\ \eqref{dotphiweak}-\eqref{dotrweak} for the weak coupling regime ($gN<20\hbar^2/M$), or Eqs.\ \eqref{dotphi1}-\eqref{dotu1} for the strong coupling regime ($gN>20\hbar^2/M$). For $gN < 20\hbar^2/M$ we used $\Omega = 2 \omega_\perp$, whereas for $gN > 20\hbar^2/M$ we used $\Omega = 2\omega_\perp/5$. In all simulations we used $g_{12}=g$. We do not see variation of $\kappa$ for different values of $\Omega$, and our results are in close agreement with Eq.\ \eqref{kappa}, plotted here with $C=4$ as a continuous line.}
\label{fig:kappa}
\end{figure}

We wish now to discuss the dependence of $\kappa$ on the interaction strength.
In Eqs.\ \eqref{Psirabi} and \eqref{Psi1} we introduced $\bar{\phi}$, the ``global'' phase difference between the two components. To a first approximation, $\bar{\phi}$ varies in time due to the energy difference between the first and second component: $\bar{\phi} = \Delta E t/\hbar$. In the limit $gN=0$, the second component is in the ground state with energy $\hbar \omega_\perp$, whereas the first has a vortex and its energy is $2\hbar \omega_\perp$, hence $\Delta E = \hbar \omega_\perp$. In general, the vortex energy in the weak-coupling limit is $\hbar\omega_{\perp} =\hbar^2/(Md^2_{\perp})$. In the TF limit, it is $\sim C\hbar^2/(MR^2)$ where $R$ is the TF radius and $C$ is an overall numerical factor that depends on the density profile and a slowly varying logarithmic factor $\log(R/\xi)$. In our model, $R^4/d^4_{\perp}$ is given by the dimensionless ratio
\begin{equation}
\frac{R^4}{d^4_{\perp}} = \frac{4}{\pi}\frac{NgM}{\hbar^2},
\end{equation}
which is small in the weak-coupling limit and large in the TF limit. A simple interpolation formula gives 
\begin{eqnarray} \label{kappa}
\nonumber \Delta E &=& \frac{\hbar^2}{M\sqrt{d^4_\perp + R^4/C^2}} = \frac{\hbar \omega_{\perp}}{\sqrt{1 + R^4/(d^4_{\perp}C^2)}}\\
&=& \frac{\hbar \omega_{\perp}}{\sqrt{1 + 4gNM/(\pi\hbar^2C^2)}}.
\end{eqnarray}

Numerically, as we show in Fig.\ \ref{fig:kappa}, we find indeed that $\kappa$ monotonically decreases with increasing interactions, and it doesn't depend pronouncedly on the Rabi frequency $\Omega$. 
Assuming that $\Delta E = \hbar \kappa$ and $C=4$, Eq.\ \eqref{kappa} is in a very good agreement with our simulations.

\begin{figure}
\centering
\vskip 0 pt
\includegraphics[width=\columnwidth,clip]{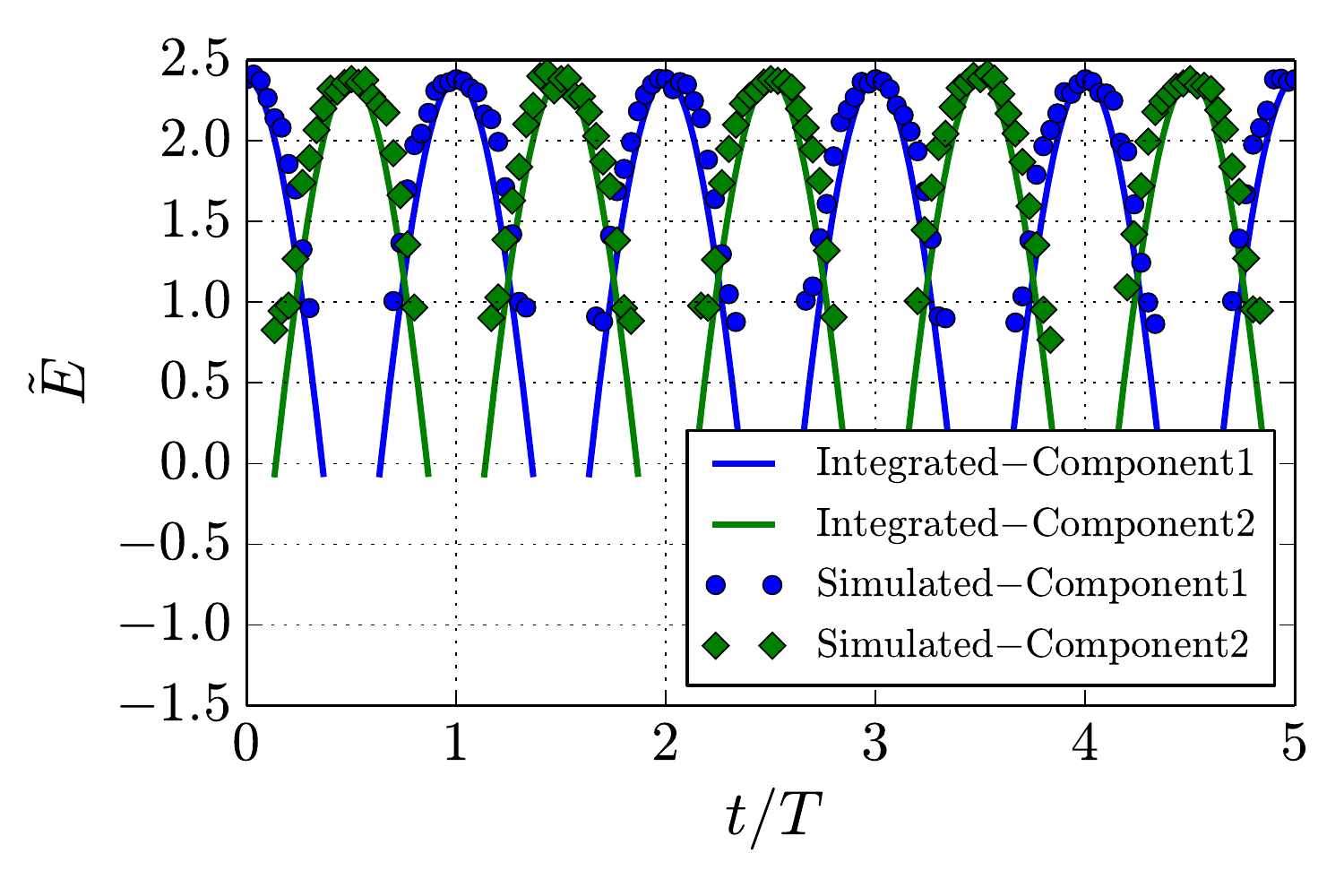}
\caption{The vortex energy $\tilde{E}$, given by Eq.\ \eqref{tildeE1}, is evaluated at the vortex core as a function of time for the simulation shown in  Fig.\ \ref{fig:single_vortex}.
 The blue and green lines depict, respectively, the vortex energy $\tilde E$ in component 1 and 2, evaluated at the solutions of  Eqs.\ \eqref{dotphi1} and \eqref{dotu1}, using $\kappa=0.485\omega_\perp$. Dots represent $\tilde E$ computed using the position of the vortex given by the GPE simulation.}
\label{fig:energy_of_single_vortex}
\end{figure}

As discussed in section \ref{coherenttrap} A, the vortex energy is time dependent, and its evolution is shown in Fig.\ \ref{fig:energy_of_single_vortex}.  Note the close agreement between the analytical expression Eq.~(\ref{tildeE1}) evaluated at the instantaneous position in each component and the corresponding simulated results.

When $g_{12}\ll g$, we observe instead a departure from the coherent behavior observed here. In analogy with the results from  Ref.\ \cite{Gall15}, we find that the system first displays incoherent features (such as delays in the vortex transfer, and trajectories which do not cross the cloud center), and for sufficiently large $g$, and small $g_{12}$, the system finally enters a regime of ``vortex trapping", where a vortex initially present inside a given component remains forever inside that same one. A sample video of incoherent dynamics, obtained with parameters as in Fig.~\ref{fig:single_vortex} but choosing this time $g_{12}=g/5$, may be found in the Supplemental Material \cite{SuppMat}.

\section{Dynamics of the coherent transfer of population}\label{JosephsonDynamics}
\begin{figure}[t]
    \centering
	\includegraphics[width=\columnwidth]{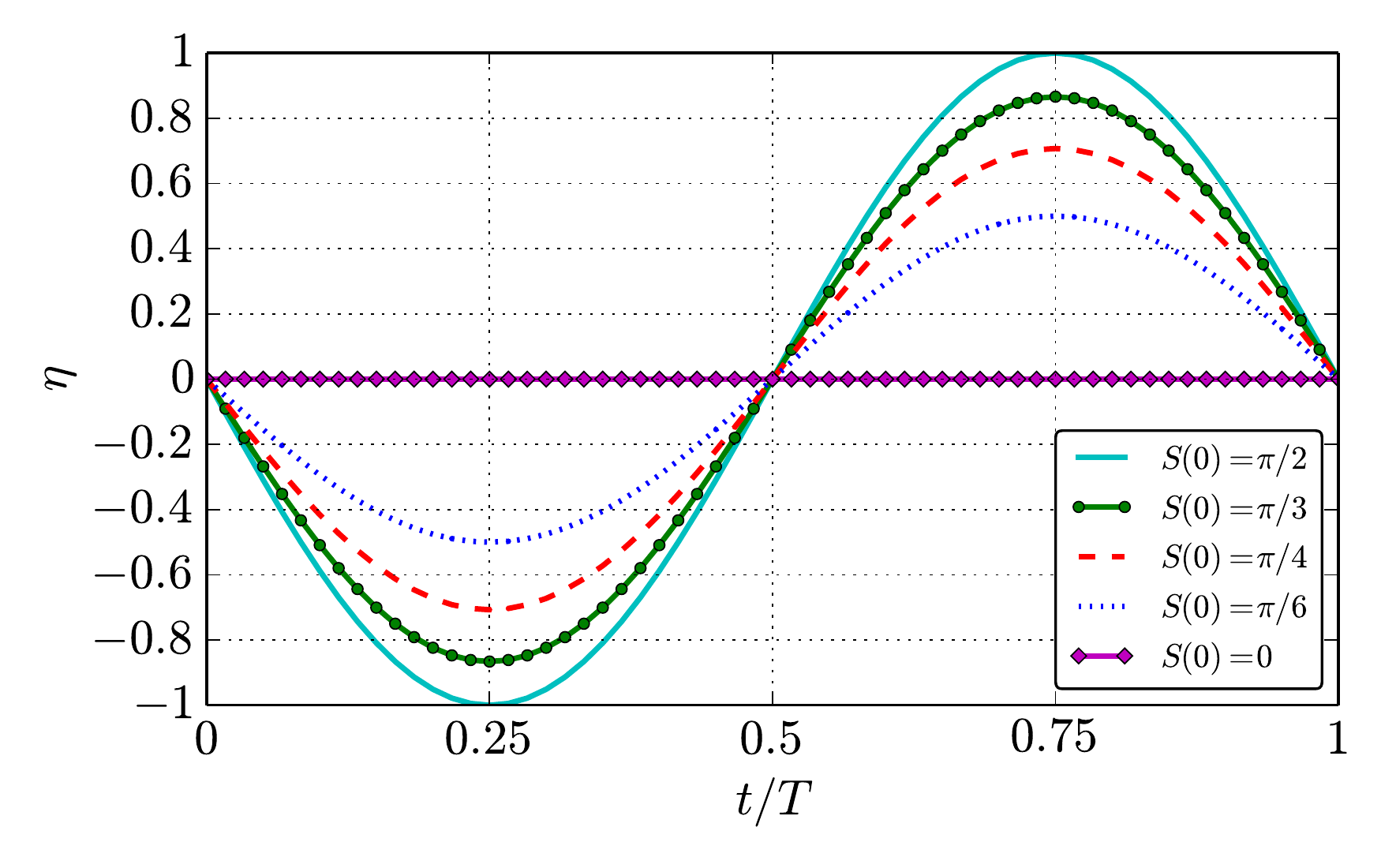}  
\includegraphics[width=\columnwidth]{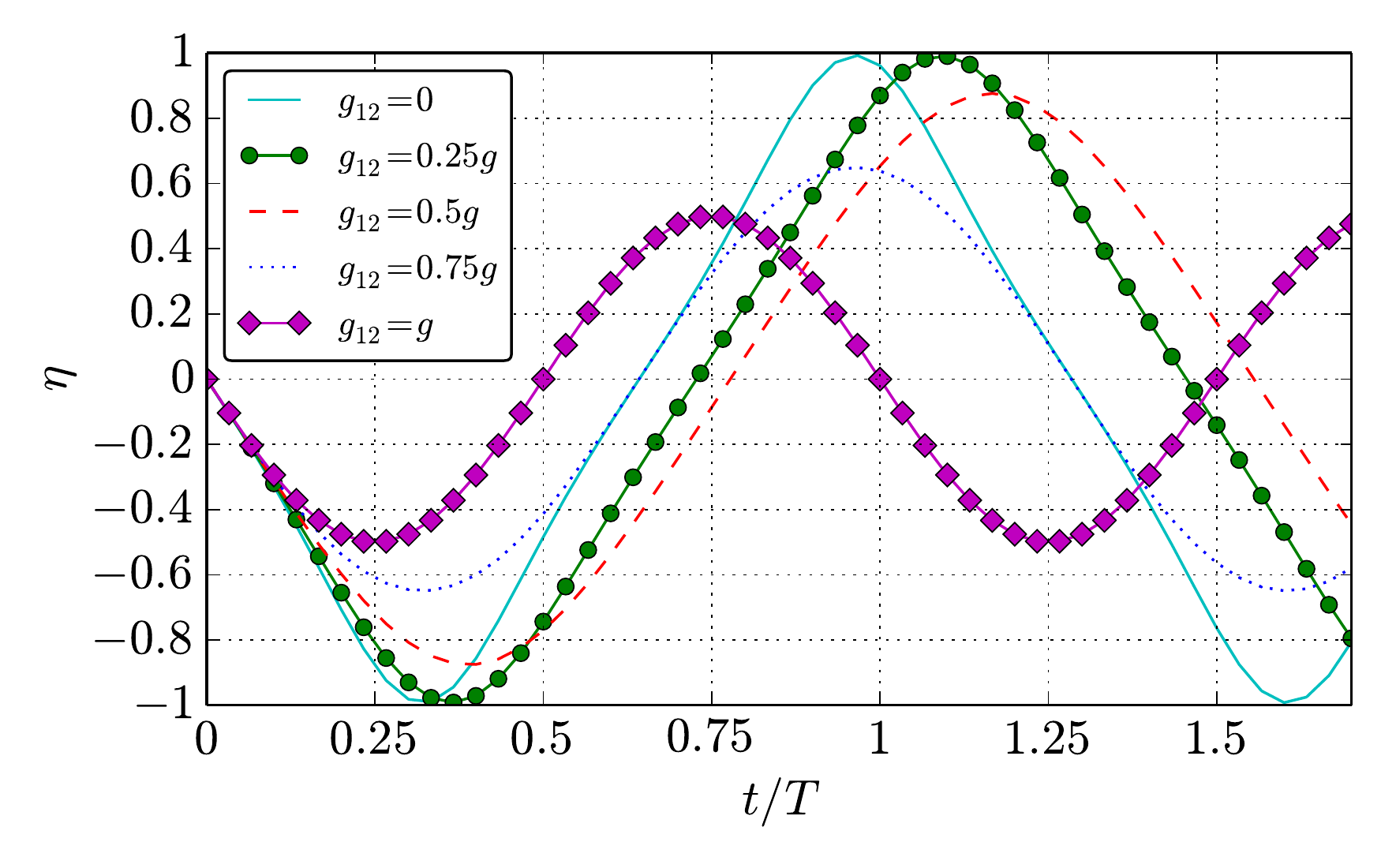}
	\caption{Pseudo-spin oscillation of Rabi-coupled BECs, obtained by the numerical solution of the coupled GP equations in a square box of area $L^2$.
	As initial conditions we take the ground state of the uncoupled BECs with $N_1 = N_2 = N/2$, $gN=40\hbar^2/M$ and $T=2\pi/\Omega=\pi M L^2/\hbar$. 
	{\bf (Top)} Dynamics of the population imbalance $\eta$ with $g = g_{12}$, for various values of the initial phase difference $S(0)$.
	{\bf (Bottom)} Evolution of the population imbalance $\eta$ at fixed phase difference $S(0)=\pi/6$, for different values of the inter-species interaction $g_{12}$;
	 the pseudo-spin oscillations become perfectly harmonic in the SU(2)-invariant case  $g_{12}=g$.}
	 \label{plot:N1-dynamics-density-interaction}
\end{figure}

We wish to study here the transfer of population (or ``pseudo-spin dynamics") induced by a coherent Rabi coupling in a two-component BEC, comparing specifically the case where both components have a uniform phase to the case where one component contains a vortex. Following Ref.~\cite{Will99}, we introduce the Ansatz for the two components' wave function
\begin{eqnarray}
\nonumber \psi_1(t,\bm r) &=& \sqrt{N_1(t)} \mathrm{e}^{iS_1(t)} \Phi_1(\bm r),\\[.2cm]
\psi_2(t,\bm r) &=& \sqrt{N_2(t)} \mathrm{e}^{iS_2(t)} \Phi_2(\bm r),
\end{eqnarray}
with real $\Phi_i(\bm r)$. Inserting this Ansatz in the coupled GP equations, one may derive Josephson-type equations of motion for the population imbalance $\eta = (N_2 - N_1)/N$ and relative phase $S = S_2 - S_1$:
\begin{eqnarray}\label{eq:GP-eta-phi}
 \frac{d}{dt} \eta &=& - k (1 - \eta^2)^{1/2} \sin(S) \equiv f(\eta, S), \\
\nonumber \frac{d}{dt} S &=& - (\mu_2 - \mu_1) + k \eta (1 - \eta^2)^{-1/2} \cos(S) \equiv g(\eta, S),
\end{eqnarray}
where $k = -\Omega \int \mathrm{d}^2 r \;\Phi_1(\bm r) \Phi_2(\bm r)$ is proportional to the Rabi frequency, and to the spatial overlap of the components.
Taking the second derivative of $\eta$ we have
\begin{equation}
\frac{d^2}{dt^2} \eta = k \frac{\eta}{\sqrt{1-\eta^2}} \dot{\eta} \sin(S) - k \cos(S) \sqrt{1-\eta^2} \dot{S}
\end{equation}
and using once more Eqs.~\eqref{eq:GP-eta-phi} we find the alternative, more transparent equation
\begin{equation}
\frac{d^2}{dt^2} \eta = - k^2 \eta + (\mu_2-\mu_1) k \sqrt{1-\eta^2} \cos(S).
\end{equation}
In uniform space, we have $\mu_2 - \mu_1 = (g-g_{12})\eta n$, with $n=N/V$. It is now not difficult to show that Eqs.~\eqref{eq:GP-eta-phi} support two kind of solutions: harmonic oscillations of the population imbalance when $(g-g_{12})\eta(0)\cos S(0) < k$, and anharmonic ones otherwise. The latter are still periodic, but display 4 changes of curvature per period; see, e.g., the thin blue line in the bottom panel of Fig.~\ref{plot:N1-dynamics-density-interaction}. Moreover, for a gas with SU(2)-invariant interactions (i.e., with $g_{12}=g$) these equations predict that the gas will perform undamped harmonic oscillations with frequency exactly equal to $\Omega$.
 This dynamics is shown in Fig.\ \ref{plot:N1-dynamics-density-interaction}, where we evolved an untrapped system in imaginary-time, reaching the ground state, and then let it evolve. Our simulations follow very closely the predictions of Eqs.~\eqref{eq:GP-eta-phi}.
Whenever $g_{12} \leq g$,  we see periodic oscillations of the population imbalance, that become harmonic and with frequency $\Omega$ for $g_{12}=g$. The amplitude of the oscillation depends on the phase difference between the two components. In particular when $g_{12}=g$ the dynamics of $\eta$ is identical to the one for a system with no interaction: the amplitude of the oscillations is proportional to $\sin S(0)$, and the period coincides with $2\pi/\Omega$, as shown in Figs.\ \ref{plot:N1-dynamics-density-interaction}.

\begin{figure}[t]
    \centering
	\includegraphics[width=\columnwidth]{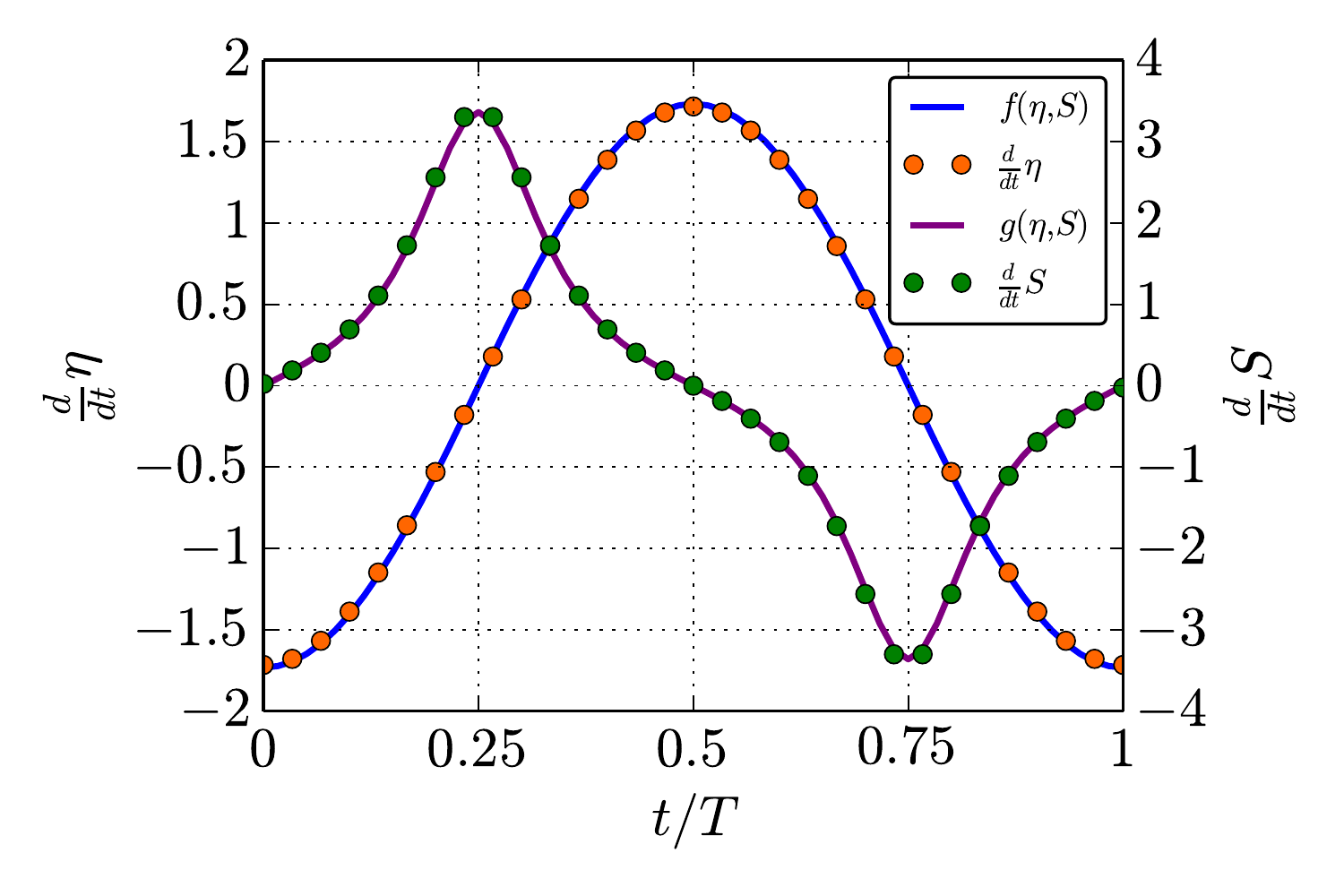}
	\caption{Evolution of population imbalance $\eta$ and relative phase $S$ for two Rabi-coupled BECs in a harmonic trap. The results of our simulations, shown with symbols, are compared with equations~(\ref{eq:GP-eta-phi}), shown as lines. As initial conditions we took the ground state of the uncoupled BECs with $N_1 = N_2 = N/2$ and $S(0) = \pi/3$. Here $gN=g_{12}N=40\hbar^2/M$ and $T=\pi/\omega$.
	} \label{plot:Comparison}
\end{figure}

Harmonic oscillations of the population difference appear also when an interacting gas is harmonically trapped. Figure~\ref{plot:Comparison} illustrates such scenario, and our simulation is seen to be in perfect agreement with Eqs.~\eqref{eq:GP-eta-phi}, with $\Phi_i(\bm r)$ the ground state wave function of the harmonic oscillator in component $i$. 

\begin{figure}
\centering
\vskip 0 pt
\includegraphics[width=\columnwidth,clip]{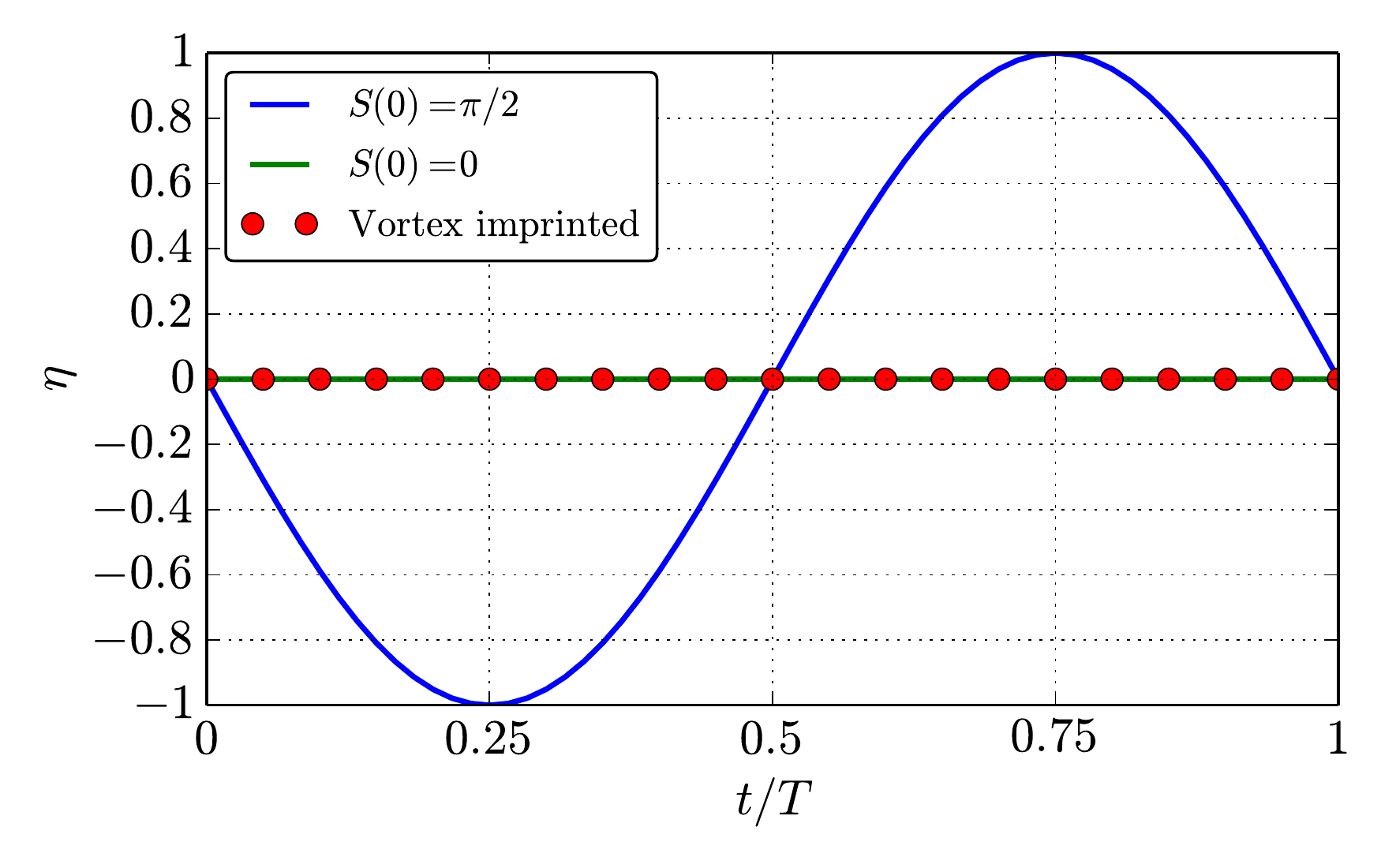}
\caption{Pseudo-spin oscillations in the presence of both trapping and interactions. The dotted line displays the population imbalance $\eta$ for the simulation shown in Fig.~\ref{fig:single_vortex}: the transfer of vorticity we observed there happens in complete absence of population transfer. For comparison, the lines show the evolution obtained starting from the same initial conditions, but without vortices in either component: green  and blue lines represent, respectively, the cases where the relative phase at $t=0$ is $0$, and $\pi/2$.}
\label{fig:pseudospinOscillationsTrappedAndInteracting}
\end{figure}

Finally we compare the case of uniform phases with the one where a single vortex is imprinted in a given trapped TF component. Our results are summarized in Fig.~\ref{fig:pseudospinOscillationsTrappedAndInteracting}. Here, we have chosen the same interaction strengths and Rabi coupling as in Fig.\ \ref{fig:single_vortex}.
 The continuous lines show the time dependence of the population of the first component when no vortex is present at $t=0$: the relative phase is in this case homogeneous across the cloud so that, in analogy with that observed previously, pronounced oscillations are observed when $S(0)=\pi/2$ (blue line); the oscillations instead disappear when $S(0)=0$, as their amplitude is  proportional to $\sin( S(0))$. The dotted line displays instead the evolution observed after phase-imprinting a vortex in the first component: the coherent transfer of vorticity observed in Fig.~\ref{fig:single_vortex} happens, to a very good approximation, in the absence of pseudo-spin oscillations. This absence may be understood by observing that, in the presence of vortices, it becomes impossible to define a global relative phase between the two components. A relative phase may still be defined locally, but the latter will evolve uniformly from 0 to $2\pi$ along a path encircling the origin, so that on one side of the trap center one will find coherent transfer of particles from component 1 to 2, while an opposite and (approximately) equal transfer will happen on the opposite side, yielding a globally vanishing pseudo-spin oscillation.

\section{Discussion and Conclusions}

The present consideration of vortex dynamics starts from classical hydrodynamics and then introduces the idea of logarithmic vortex interactions. In contrast, the time-dependent variational Lagrangian formalism focuses on the phase $S$ of the condensate wave function for various components, and it is worth examining this aspect in more detail.

The simplified Lagrangian density in (\ref{Lsimp}) studies the phase of each condensate $S_1$ and $S_2$, omitting any spatial and temporal variation in the densities $n_1$ and $n_2$ along with the trap potential.  The time derivatives yield the time-dependent term $T$ in the Lagrangian,  $L = T-E$, and the gradient terms yield the kinetic term in the GP energy $E_{\rm GP}$.  In the limit of relatively weak Rabi rf coupling ($l_\Omega \gtrsim r_{12}$), it is natural to assume that the phase functions are those of the pure vortex with $S_j[\bm r-\bm r_j(t)]$ from Eq.~(\ref{phasej}).  As a result, the time derivative becomes  $\dot{S}_j= -\dot{\bm r}_j\cdot \bm \nabla S_j$.  Thus  $\bm \nabla S_j$ determines both the time term $T$ and the kinetic energy part of $E$ in the Lagrangian $L=T-E$.

The presence of Rabi rf coupling alters this picture because it provides an additional term $E_\Omega$ in the energy.  This energy gives rise to an additional force $\bm F_\Omega=-\bm\nabla E_\Omega$ with different dependence on  the vortex separation and even a different sign from that arising from the intervortex potential.   Indeed, for two positive vortices, one in each component of a two-component  unbounded condensate, the rotation rate is in the negative sense.  In contrast, two  positive vortices in classical hydrodynamics or in a one-component BEC would rotate in the positive sense.

The  picture changes significantly for strong Rabi coupling $l_\Omega\lesssim r_{ij}$ for uniform condensates (or $l_\Omega\lesssim R$ for TF trapped condensates).    The coupling energy now varies linearly with the vortex separation, and the rotation frequency in (\ref{rot2}) agrees with that found with somewhat different methods by the Trento group~\cite{Tylu16}.

The results of our numerical simulations closely match the theoretical derivations. In particular, we have verified that the rotation frequency of a pair of positive vortices changes sign as a function of the applied Rabi frequency, and our results converge to the theoretical prediction of the Lagrangian formulation at strong Rabi coupling, and to the value expected for single off-centered vortices in a cylindrical container in the opposite limit of weak Rabi coupling. 

Moreover, we have verified that if a single vortex is imprinted in only one of the components, the Rabi coupling drives an interesting dynamics, where the vortex is coherently transferred from one component to the other. Finally, we have shown that this coherent transfer of vorticity  happens with no transfer of population.

The numerical results rely on an open source library~\cite{TSCode}, and we wish to promote the practice of open science by making the steps involved in the simulations available online~\cite{TSNotebook}. This will ensure a straightforward reproduction of the plots, and we also hope that it will make it easier to extend our work.

\section*{Acknowledgements} ALF is grateful to S.~Stringari for bringing this problem to his attention during an INT  workshop in March 2015 and  thanks the Institute for Nuclear Theory at the University of Washington for its hospitality and the Department of Energy for partial support during the inception of this work.  A visit to the University of Otago, Dunedin, N.\ Z.\ with A.\ Bradley and colleagues allowed ALF to develop some of these ideas.
The Aspen Center for Physics and the NSF Grant \#1066293 provided a supporting  environment where ALF developed a talk on this subject and initiated this collaboration with PM.
PM acknowledges funding from a ``Ram\'on y Cajal" fellowship.
PW acknowledges financial support from the ERC (Consolidator Grant QITBOX), and the computational resources granted by the High Performance Computing Center North (SNIC 2015/1-162).
PM and PW further acknowledge support from MINECO (Severo Ochoa grant SEV-2015-0522 and FOQUS FIS2013-46768), Generalitat de Catalunya (SGR 874 and 875), and the Fundaci\'o Privada Cellex.
LC acknowledges the Center of Studies and Activities for Space (CISAS) ``Giuseppe Colombo" for financial support.

\section*{Appendix}

This Appendix provides an analytical derivation of the  results given in Eqs.~(\ref{E+}) and (\ref{E-}).   It involves   Landen transformations  for complete elliptic integrals.
For the $q_1q_2 = +$ case, a change of variable $\theta=\pi/2-\chi$ and symmetry yield the relevant dimensionless integral
\begin{eqnarray}
{\cal I}^+ &=& 4\int_0^{u_0} u\,du \int_0^{\pi/2} d\chi \left[ 1-\frac{u^2-1}{u^2+1}\,\frac{1}{\sqrt{1-k^2\sin^2 \chi}}\right] \nonumber\\
 &=&  4\int_0^{u_0} u\,du \left[ \frac{\pi}{2}-\frac{u^2-1}{u^2+1}K(k)\right] ,
\end{eqnarray}
where $u_0 = 2\Lambda/r_{12}$.  Here $K(k)$ is the complete elliptic integral of the first kind and $k = 2u/(1+u^2)$ [note that $K(k)$  is the usual notation in mathematics, but it differs from EllipticK($k^2$) in Mathematica, which uses the variable $k^2$ instead of $k$].

Elliptic integrals obey identities known as Landen transformations~\cite{dlmf}. For a given $k<1$, they involve the sequence of transformations $k'= \sqrt{1-k^2}$ followed by $k_1=(1-k')/(1+k')$. For example, if $u^2<1$,
\begin{equation}
K\left(\frac{2u}{1+u^2} \right)= (1+u^2)K(u^2),
\end{equation}
whereas if $u^2>1$,
\begin{equation}
K\left(\frac{2u}{1+u^2} \right)= (1+u^{-2})K(u^{-2}),
\end{equation}
These results
 allow the previous integral to be rewritten in a different form (it is necessary to separate the two regions $u<1$ and $u>1$, and the latter contains the cutoff) ${\cal I}^+ = {\cal I}_<^+ + {\cal I}_>^+$.

The first integral is straightforward (change variable to  $v = u^2$) and is  simply a number
\begin{equation}
{\cal I}_< ^+= 2\int_0^1 dv\left[\frac{\pi}{2} +K(v) -vK(v)\right]  = \pi +4G-2,
\end{equation}
where $G\approx 0.91597$ is Catalan's constant~\cite{GR}.
For the second integral,  the new variable $v=u^{-2}$ gives
\begin{equation}
{\cal I}_>^+= 2\int_{u_0^{-2}}^1 dv \left[\frac{\pi}{2v^2}-\frac{K(v)}{v^2} + \frac{K(v)}{v}\right].
\end{equation}
In the limit $u_0\to \infty$, the first two terms give a  convergent integral $\int_0^1\,dv\,v^{-2}\left[\pi-2K(v)\right]= -\pi+2$.  The last term can be written
\begin{equation}
\nonumber \approx 2\int_0^1\frac{dv}{v}\left[K(v)-\frac{\pi}{2}\right] + 2\pi \ln u_0= 2\pi\ln 2-4G +2\pi \ln u_0.
\end{equation}
The sum of all the various terms gives the final result ${\cal I}^+\approx 2\pi\ln (2u_0) \approx 2\pi\ln(4\Lambda/r_{12})$,  leading to Eq.~(\ref{E+}).

For the other integral ${\cal I}^-$,  the same change of variable $\theta=\pi/2-\chi $ gives
\begin{equation}
\nonumber {\cal I}^- = 4\int_0^{u_0} u\,du \int_0^{\pi/2} d\chi \left[ 1-\frac{1}{u^2+1}\,\frac{1+u^2-2u^2\sin^2\chi}{\sqrt{1-k^2\sin^2 \chi}}\right] .
\end{equation}
A bit of algebra shows that
\begin{equation}
{\cal I}^{-} = 2\int_0^{u_0} u\,du \left[(1-u^2)E(k) + (1+ u^2)K(k)\right] ,
\end{equation}
where, as before, $u_0 =2\Lambda/r_{12}$,  $k = 2u/(u^2+1)$, and $E(k)$ is the complete elliptic integral of the second kind.

The Landen transformation~\cite{dlmf} now gives $E(k) = (1+k')E(k_1)-k'K(k)$. In particular, if $u^2<1$,  \begin{equation}
2\left[(1-u^2)E(k) + (1+u^2)K(k)\right] = 4E(u^2).
\end{equation}
In contrast, if $u^2>1$, a similar analysis gives
\begin{multline}
2\left[(1-u^2)E(k) + (1+u^2)K(k)\right] =\\ 
=  4\left[u^2E\left(\frac{1}{u^2}\right)+ \frac{1-u^4}{u^2}K\left(\frac{1}{u^2}\right)\right].
\end{multline}
Hence ${\cal I}^- ={\cal I}_<^- +{\cal I}_>^-$, and the second piece again contains the divergent logarithmic  part.

The first integral is a known quantity, and the variable $v = u^2$ gives
\begin{equation}
{\cal I}_<^- = 2\int_0^1\,dv\,E(v) = 1+ 2G \approx 2.83193.
\end{equation}
For the second integral, the substitution $v = 1/u^2$ yields  a logarithmic  divergence near the origin. An expansion of the integrand for small $v$ gives the approximate behavior that can be added and subtracted.  In this way,
\begin{equation}
\nonumber {\cal I}_>^- \approx  2\int_0^1 \frac{dv}{v^3}\left[E(v) + (v^2-1)K(v) -\frac{\pi v^2}{4} \right] + \pi \ln(u_0).
\end{equation}
Here, the first integral is finite, and the second term is the logarithmic leading contribution $\pi \ln(2\Lambda/r_{12})$.  Numerical integration gives
$
{\cal I}_>^- \approx 0.131053 + \pi \ln(2\Lambda/r_{12}).
$
The sum of these various terms yields
\begin{equation}
\nonumber
{\cal I}^-
\approx \pi \ln(2\Lambda/r_{12})+ 2.96298
 \approx \pi \ln(5.1361\Lambda/r_{12}),
 \end{equation}
 which is the value quoted in Eq.~(\ref{E-}).

\widetext
\begin{center}
\textbf{\large Supplemental Material:
			Vortex dynamics in coherently coupled Bose-Einstein condensates}\\
\vspace{4mm}
{Luca Calderaro,$^1$ Alexander L. Fetter,$^2$ Pietro Massignan,$^3$ and Peter Wittek$^{3,4}$}\\
\vspace{2mm}
{\em \small
$^1$Dipartimento di Ingegneria dell'Informazione, Universit\`a di Padova, 35131 Padova, Italy\\
$^2$Departments of Physics and Applied Physics, Stanford University, Stanford, CA 94305-4045, USA\\
$^3$ICFO -- Institut de Ciencies Fotoniques, The Barcelona Institute of Science and Technology, 08860 Castelldefels (Barcelona), Spain\\
$^4$University of Bor{\aa}s, 50190 Bor{\aa}s, Sweden\\
}\end{center}

See \url{https://www.youtube.com/watch?v=hKxT-qv4oHI} for a video displaying a complete simulation of dynamics of two counter-rotating vortices, one in each component ($q_1=-q_2=1$), with $l_\Omega/r_{12}=0.5$ and $r_{12}/\xi=10$. The unit of length in this simulation is $r_{12}$, the distance between the two vortices at $t=0$.

Moreover, see \url{https://www.youtube.com/watch?v=LGyrvHOQUDk} for a video displaying a complete simulation of coherent evolution of a single vortex in a two-component condensate with $g_{12}=g$, and  \url{https://www.youtube.com/watch?v=ebD7LUmLRJ4} for a video of incoherent evolution obtained with a smaller inter-component interaction, $g_{12}=g/5$. In both these videos, all other parameters are set as in Fig.\ \ref{fig:single_vortex}.
 
\end{document}